\documentclass[fleqn,10pt]{wlscirep}

\usepackage[utf8]{inputenc}
\usepackage[T1]{fontenc}

    \usepackage{xcolor}
    \usepackage{gensymb}
    \usepackage{svg}
    \usepackage{graphicx}
    \usepackage{hyperref}
    \usepackage{dcolumn}
    \usepackage{bm}

    \usepackage{soul}
\usepackage{chngcntr}

    \newcommand{\kvec}{\textbf{k}}
    \newcommand{\micron}{\mu\textrm{m}}
    
    \newcommand{\metastable}{2^{3\!}S_1}%
    \newcommand{\mhe}{\textrm{He}$^*~$}%

    \newcommand{\op}[1]{\widehat{#1}}
    \newcommand{\dagop}[1]{\widehat{#1}^{\dagger}}
    \newcommand{\bo}[1]{{\mathbf{#1}}}
    \newcommand{\mc}[1]{{\mathcal{#1}}}
    \newcommand{\wt}[1]{{\widetilde{#1}}}
    \newcommand{\wb}[1]{{\overline{#1}}}
    
    \newcommand{\matri}[2]{\left[\begin{array}{#1}#2\end{array}\right]}
    \newcommand{\arr}[2]{\begin{array}{#1}#2\end{array}}
    \newcommand{\ifmath}[2]{\left\{\begin{array}{#1}#2\end{array}\right.}
    \newcommand{\real}[1]{{\rm Re}\left[{#1}\right]}

    \newcommand{\eqn}[1]{(\ref{#1})}
    \newcommand{\eq}[2]{\begin{equation}\label{#1}#2\end{equation}}
    
    \newcommand{\leqs}[2]{\begin{subequations}\label{#1}\begin{eqnarray}#2\end{eqnarray}\end{subequations}}
    \newcommand{\eqs}[2]{\begin{subequations}\label{#1}\begin{eqnarray}#2\end{eqnarray}\end{subequations}}
    \newcommand{\eqa}[2]{\begin{eqnarray}\label{#1}#2\end{eqnarray}}

\begin{document}

\title{On the survival of the quantum depletion of a condensate after release from a magnetic trap}

\author[1]{J. A. Ross}
\author[2]{P. Deuar}
\author[1]{D. K. Shin}
\author[1]{K. F. Thomas}
\author[1]{B. M. Henson}
\author[1]{S. S. Hodgman}
\author[1*]{A. G. Truscott}
\affil[1]{Research School of Physics, Australian National University, Canberra 0200, Australia}
\affil[2]{Institute of Physics, Polish Academy of Sciences, Aleja Lotnik\'{o}w 32/46, 02-688 Warsaw, Poland}

\affil[*]{andrew.truscott@anu.edu.au}

\begin{abstract}

We present observations of the {high momentum tail} in expanding Bose-Einstein condensates of metastable Helium atoms released from a harmonic trap. 
The far-field density profile exhibits features that support identification of the tails of the momentum distribution as originating in the in-situ quantum depletion prior to release. 
Thus, we corroborate recent observations of slowly-decaying tails in the far-field beyond the thermal component. 
This {observation} is in conflict with the hydrodynamic theory, which predicts that the in-situ depletion does not survive when atoms are released from a trap.  
Indeed, the depleted tails even appear stronger in the far-field than expected before release, and we discuss the challenges of interpreting this in terms of the Tan contact in the trapped gas.
In complement to  {these observations, full quantum} simulations of {the} experiment show that, under the right conditions, the depletion can persist into the far field after expansion.
Moreover, the simulations provide mechanisms for survival and for the the large-momentum tails to appear stronger after expansion due to an acceleration of the depleted atoms by the mean-field potential.  
However, while in qualitative agreement, the final depletion observed in the experiment is much larger than in the simulation. 
\end{abstract}

\flushbottom
\maketitle
\thispagestyle{empty}

\section*{Introduction}

	{In the Bogoliubov description of an ultracold interacting superfluid, the ground state}  
	is composed of a macroscopically-occupied condensate and correlated particle pairs due to s-wave interactions between constituent particles \cite{Bogolubov47,Vogels02}.
	{A consequence of these pairs is  
	that excited} single-particle modes 
	are populated even at zero temperature. 
	This is the \emph{quantum depletion} of the condensate and presents as an occupation of single particle modes, which at large momentum $p$ decays \cite{PitaevskiiStringari,Decamp18} like $p^{-4}$.

	Since the realization of atomic Bose-Einstein condensates (BECs) there has been considerable experimental \cite{Stewart10,Wild12,Chang16,Makotyn14,Eigen18,Xu06,Vogels02,pieczarka20,Lopes17_depletion,Cayla20,Kuhnle11,Sagi12,Fletcher17,Lopes17_quasiparticle,Mukherjee19,Carcy19} and 
	theoretical \cite{Colussi20,Kira15_coherent,Decamp18,Smith14,Qu16,Braaten10,Braaten11,Rakhimov20,Braaten08,Zhang09,Combescot09,Werner12_boson,Werner12_fermion,Sinatra00,Deuar11} interest in the 
	Bogoliubov theory \cite{Vogels02,Steinhauer03,Lopes17_quasiparticle,Sinatra00,Deuar11} (and 
	quantum depletion specifically \cite{Lopes17_depletion,Chang16,Xu06,pieczarka20,Cayla20,Cayla22}).
	In contrast to the case of liquid helium, where the depleted fraction is large (of order 93\% of the fluid \cite{Dmowski17,Glyde00,Moroni04}) due to the strong interparticle interactions, the depletion is generally very small (less than 1\% \cite{Lopes17_depletion,Chang16}) in weakly-interacting dilute gases.
	The intimately related thermodynamic {(Tan's)} contact 
	has also received growing attention, \cite{Stewart10,Tan08_momentum,Tan08_energetics,Tan08_virial, Braaten10,Braaten11,Colussi20,Makotyn14,Eigen18,Decamp18,Smith14,Chang16,Qu16,Wild12,Hoinka15,Rakhimov20,Braaten08,Smith14,Kuhnle11,Sagi12,Fletcher17,Mukherjee19,Carcy19,Zhang09,Combescot09,Werner12_boson,Werner12_fermion,Cayla22}, in part due to Tan's proof that the contact is directly related to the amplitude of the $p^{-4}$ tail \cite{Tan08_momentum}.
	
	{Experiments examining}
	the large-momentum tails  {have typically employed} Feshbach resonances to enhance interactions in ultracold gases and produce a depleted fraction visible in the far-field with standard imaging techniques, but the power-law tails have proven elusive \cite{Makotyn14,Eigen18} in this regime. 
	A handful of theories have emerged \cite{Kira15_coherent,Colussi20,Smith14} which elucidate the role played by many-body interactions {in modifying the momentum distribution during}  
	the evolution following a quench to a large scattering length. 
	A very recent experiment \cite{Tenart21} was able to detect pairs of atoms with anticorrelated momenta in the far-field by use of an optical lattice to create a BEC in a high-density, strongly-interacting regime.
	However, measurements in the weakly-interacting regime have returned unexpected results.
	A previous experiment reported the presence of power-law-like tails in the far-field  distribution after releasing a BEC of metastable helium from a harmonic optical trap \cite{Chang16}.
	This was surprising because conventional wisdom argues that the density decreases adiabatically during expansion (even when the trap release is non-adiabatic), motivating a hydrodynamic approximation wherein the tails are predicted to vanish \cite{Qu16,Xu06}. 
	Moreover, the tails were reported to be approximately six times heavier than predicted by Bogoliubov theory.
	It is important to verify the anomaly and understand its origin because far-field measurements play a central role in the study of ultracold gases. 
	The prospect of extracting correlated depleted pairs from a zero-temperature ground state is also conceptually, and possibly technologically, interesting in itself. 

	To these ends, we measure the far-field momentum distribution of a BEC of metastable helium (\mhe) expanding from a harmonic trap. 
	We observe tails in the large-momentum part of the (far-field) condensate wavefunction whose amplitude depends nonlinearly on the condensate population, and whose density profile is consistent with a $p^{-4}$-like power law decay, in a manner consistent with the predictions of the Tan and Bogoliubov theory.
	Specifically, the amplitude of the far-field momentum tails is shown to have a linear relationship with the product of the condensate population and peak density, as predicted by both theories.
	However, there is a quantitative difference in amplitude between the predicted and measured values. 
	Our measurements are complemented by numerical simulations of the dynamics of the momentum distribution after the trap release using a Stochastic Time-Adaptive Bogoliubov (STAB) method in the positive-P framework \cite{Deuar11,Kheruntsyan12}. 
	{These demonstrate a mechanism for survival associated with the non-adiabatic release of the trap, and suggest 
	that the depleted particles acquire additional kinetic energy from the mean-field energy of the condensate during the subsequent adiabatic expansion.}
	These factors result in an amplification of the density of the far-field momentum tails relative to the in-situ values {by a factor of up to about two}, and are absent from the hydrodynamic approximation. 
	However, even taking these effects into account, the amplitude of the measured tails is still significantly larger than expected from the simulations.

\subsection*{Quantum depletion of the condensate by contact interactions}

	Before presenting our results, let us introduce the central theoretical assumptions and predictions relevant for this work. The Hamiltonian of a homogeneous system of interacting bosons can be written in terms of plane-wave field operators $\hat{a_\kvec}$, labeled by the wavevector $\kvec=\textbf{p}/\hbar$, and diagonalized by the Bogoliubov transformation to a free Bose gas of collective excitations through the operator transformation $\hat{b}_{\kvec}^\dagger = u_k \hat{a}_\kvec^\dagger + v_k \hat{a}_{-\kvec}$ \cite{Bogolubov47,PethickSmith}. The collective excitations are superpositions of particles with opposite momenta \cite{Vogels02}, and the $u_k$ and $v_k$ coefficients are given by 
	\begin{align}
	u_k &= \cosh \theta_k, \qquad v_k = \sinh\theta_k\\
	\theta_k &= \frac{1}{2}\log\frac{\hbar^2k^2/2m}{\epsilon(k)} <0
	\end{align}
	where the denominator is the quasiparticle dispersion
	\begin{equation}
		\epsilon(k) = \sqrt{\left(\frac{\hbar^2k^2}{2m}\right)^2 + gn\frac{ \hbar^2k^2}{m}}.
		\label{bogfreq}
	\end{equation}
	determined by the particle density $n$, the atomic mass $m$, and the effective interaction strength $g=4\pi\hbar^2a/m$, where $a$ is the s-wave scattering length \cite{PitaevskiiStringari,PethickSmith}.
	In the non-interacting ($a\rightarrow0$) limit, $u_k=1$ and $v_k=0$, so the transformation reduces to the identity and the dispersion is that of free particles. 
	The occupation of single-particle momentum modes can be found using the inverse transformation and is given by
	 \begin{align}
	 \rho(\kvec) &= \langle\hat{a}_\kvec^\dagger\hat{a}_\kvec\rangle\\
		 &=\left(u_{k}^{2}+v_{k}^{2}\right)\langle \hat{b}_{\kvec}^{\dagger}\hat{b}_{\kvec}\rangle + v_{k}^{2},
		 \label{eqn:popstats}
	 \end{align}
	wherein the quasiparticle population statistics follow the canonical ensemble as \cite{PitaevskiiStringari,Chang16} $\langle \hat{b}^\dagger_\kvec\hat{b}_\kvec\rangle = (\exp[\epsilon(k)/k_B T]-1)^{-1}$. At finite temperatures, quasiparticle modes are thermally populated and deplete the condensate.  Even at zero temperature, when the thermal fraction vanishes, the $v_k^2$ term in Eqn (\ref{eqn:popstats}) persists giving a zero-temperature population of excited 
	{particles \cite{Olshanii03,Decamp18,Chang16}} which decays as \cite{PethickSmith,PitaevskiiStringari,Chang16}  $\lim_{k\rightarrow\infty}\rho(\kvec)\propto k^{-4}$. 
	Bogoliubov's theory makes accurate predictions of the total depleted population in ultracold atomic Bose-Einstein condensates (BECs) \cite{Xu06,Lopes17_depletion} and exciton-polariton condensates in solid substrates \cite{pieczarka20}.
	
    In the case of a harmonically trapped gas, one can employ the local-density approximation (LDA) to compute the amplitude of the $k^{-4}$ tail by  integrating $v_k^2$ across a Thomas-Fermi distribution \cite{Chang16}. 
    One can also compute the expected amplitude of the tails using thermodynamic relations between the condensate mean-field energy and  momentum distribution: 
    The amplitude of the tails was shown by Tan to be exactly the quantity called the \emph{contact}, which is proportional to the derivative of the energy with respect to the s-wave scattering length \cite{Tan08_momentum, Braaten11}.
	For a Bose gas at equilibrium in a harmonic trap, the tail amplitude can be calculated using Tan's original theorems. The two-body \emph{contact intensity} is defined by \cite{Tan08_momentum,Braaten11}
	\begin{equation}
		C = \lim_{k\rightarrow\infty}k^4\rho(k),
		\label{eqn:MomentumDef}
	\end{equation}
	which is related to the total contact (or just \emph{contact})  $\mathcal{C} = \int C(\textbf{r}) d^3 \textbf{r}$.
	The contact can be derived from the total energy $E$ through the \emph{adiabatic sweep theorem} \cite{Tan08_energetics},
	\begin{equation}
		\mathcal{C} = \frac{8\pi m a^2}{\hbar^2}\frac{\partial E}{\partial a}.
		\label{eqn:sweep_theorem}
	\end{equation}
	Applying this to the Thomas-Fermi energy of a harmonically trapped condensate,
		\begin{equation}
		\frac{E}{N_0} = \frac{5}{7}\mu = \frac{5}{7} \frac{\hbar \bar{\omega}}{2} \left(\frac{15 N_0 a}{a_\textrm{HO}}\right)^{2/5},
		\label{mu}
	\end{equation}
	where $a_\textrm{HO} = \sqrt{\hbar/(m \bar{\omega})}$ is the harmonic oscillator length and $\bar{\omega}=\sqrt[\uproot{2}\scriptstyle 3]{\omega_x \omega_y \omega_z}$ is the geometric trapping frequency \cite{PitaevskiiStringari,PethickSmith}, leads to the expression
	\begin{equation}
		\mathcal{C} = \frac{8\pi}{7} \left(15^{2}(a N_0)^{7} \left(\frac{m \bar{\omega}}{\hbar}\right)^{6}\right)^{1/5},
		\label{eqn:TotalHarmonicContact}
	\end{equation}
	for the contact and thus the asymptotic momentum (density) distribution $n(k)$ of the in-situ condensate is,
	\begin{equation}
		\lim_{k\rightarrow\infty} n(k) = {\frac{\mathcal{C}}{k^4}} = \frac{64\pi^2a^2}{7} \frac{N_0n_0}{k^4},
		\label{eqn:pred_scaling}
	\end{equation}
	which depends on the peak density of the harmonically trapped condensate, in turn given by
	\begin{equation}
		n_0 = \frac{1}{8 \pi}\left( (15N_0)^2 \left(\frac{m \bar{\omega}}{\hbar\sqrt{a}}\right)	 ^{6}\right)^{1/5}.
		\label{eqn:n0}
	\end{equation}
	{Note that hereon we refer to the momentum distribution $n(\kvec)$ rather than the occupation numbers $\rho(\kvec) = n(k) d^3\kvec/(2\pi)^3$, and that } {the total number of atoms in this normalisation is $N=\frac{1}{(2\pi)^3}\int d^3 \kvec\, n(\kvec)$.}

\section*{Results}

\subsection*{Experimental measurements}

    Our experimental sequence began with BECs consisting of between $2\times 10^5$ and $5\times 10^5$ $^4$He$^*$ atoms, spin-polarized in the $\metastable(m_J=1)$ state and cooled to $\sim$ 300 nK by forced evaporative cooling in a harmonic magnetic trap generated by field coils in a Bi-planar Quadrupole Ioffe configuration \cite{Dall07}. 
    {The trap was then switched off with a $1/e$ time of ${\tau_{\rm release}}\approx38\mu$s. The condensates were allowed to expand for 2ms before we transferred about one quarter of the initial $m_J=1$ condensate into the magnetically insensitive $m_J=0$ state with a radio-frequency (RF) Landau-Zener sweep to  preserve it against distortion by stray magnetic fields during the free fall to the detector}.  
	We deflected the $m_J=\pm 1$ clouds away from the detector with a Stern-Gerlach scheme immediately after the RF pulse by switching on a magnetic field.
	The centre of mass of the cloud then impacts on the detector after a $\tau = 417$ms time of flight following the trap switch-off.

	{
    Investigations of the quantum depletion in \mhe are challenging because the absence of a known Feshbach resonance precludes control over the contact $\mathcal{C}\propto((a N_0)^7\bar{\omega}^6)^{1/5}$ via the scattering length $a$. 
	Given the small fixed $a=7.512$nm \cite{Moal06}, we test the validity of Eqn. (\ref{eqn:pred_scaling}) for describing 
	the far-field by varying the density of the gas, $n\propto\left(N_{0}\bar{\omega}^3\right)^{2/5}$ (c.f. Eqn (\ref{eqn:pred_scaling})). 
	To achieve this we used two trap configurations with $(\omega_x,\omega_y,\omega_z)\approx 2\pi(45,425,425)$ Hz (geometric mean $\bar{\omega} = 2\pi \cdot201$ Hz) and 
	$\approx2\pi (71,902,895)$ Hz ($\bar{\omega} = 2\pi \cdot393$ Hz) where  the frequencies are known within 1\% the (weak) axis of symmetry is horizontal.
	We varied the endpoint of the evaporative cooling ramp to adjust the number of atoms in the condensate. }

Our experiment uses single-particle detection with multichannel-plate and delay-line detector (MCP-DLD) stacks \cite{Manning10} after a long time of flight (hence in the far-field regime) enabled by the large (19.8eV) internal energy \cite{Hodgman09} of the  metastable $\metastable$ state, 
He$^*$. 
    {The unique capabilities of such setups have}
    permitted the observation of many-body momentum correlations \cite{Hodgman11,Dall13} and the Hanbury Brown-Twiss effect in both condensed \cite{Schellekens05,Jeltes07,Manning10,Dall11,Perrin07,Perrin12} and quantum depleted atoms \cite{Cayla20,Cayla22}. 
    We are thus able to reconstruct the full single-atom momentum distribution in three dimensions and examine the dilute far-field momentum tails of the $m_J=0$ clouds in detail.

\begin{figure}[htb]
	\centering
	        \includegraphics[width=0.7\textwidth]{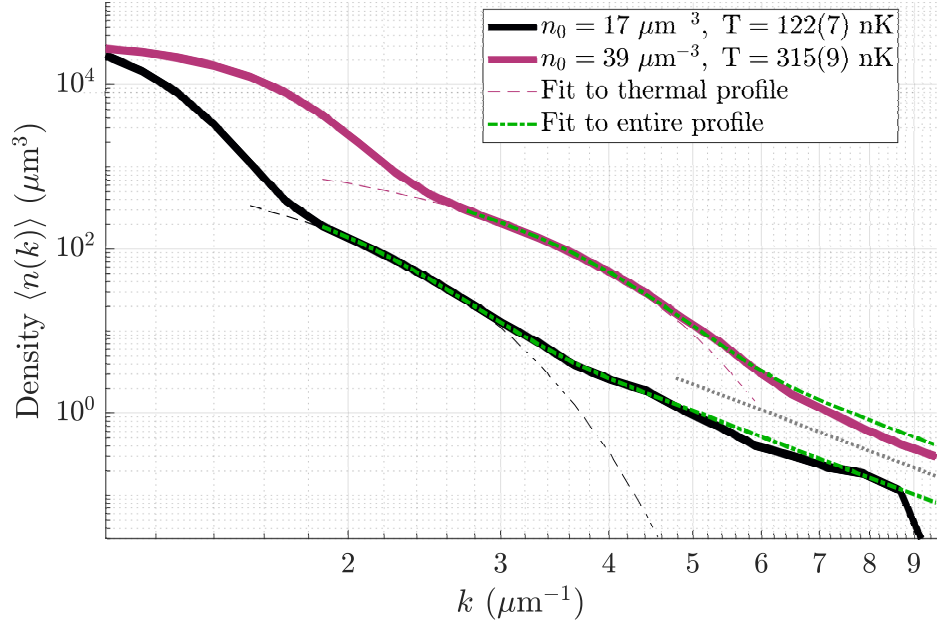}
	        \caption{The measured far-field density of particle momenta from two trap configurations (black and magenta). Three regions are shown: At low $k$ the parabolic distribution of the BEC dominates. For larger $k$, the thermal parts (fits shown by dashed lines) decay super-exponentially as $e^{-k^2}$. For even larger $k$, these give way to the {surmised quantum} depletion region. A combined fit of the form $n_T(k) + C_4/k^4$ (green dot-dash lines) yields temperatures consistent with the thermal fit and also an amplitude $C_4$ of the depleted tail. The grey dotted line is a guide to the eye showing a $k^{-4}$ decay. Due to constraints of the detector geometry (see supplementary materials for details), these profiles were integrated over two spherical segments, each subtending an angle of $\pi/6$ radians with the $\pm z$ axes (see also Figure \ref{fig:sequence}). The detector shows signs of saturation for low $k (\lesssim 1.5~\micron^{-1}$). These factors imply the total area under the curves is less than the average number of trapped atoms. }
	        \label{fig:nk_density}
	\end{figure}

	In {Figure \ref{fig:nk_density}} 
	we show the empirical far-field density $n(k)$ for two data collection runs at the extreme values of $n_0$ we used. 
	The black (purple) correspond to condensates with an average of $3.5\times10^5 (4.5\times10^5)$ atoms and a thermal fraction of 9\% (10\%).
	The (geometric) trap frequencies were $2\pi\cdot$201 and $2\pi\cdot 393$ Hz, and the healing length $\xi = \hbar/\sqrt{2mgn_0}$ at the center of these clouds were $56~\micron$ and $36~\micron$, respectively.
	The three regimes of the condensate, thermal depletion, and quantum depletion span over five orders of magnitude in density. The thermal part of the distribution is well fitted by the momentum distribution of an ideal Bose gas \cite{Dalfovo99}
	\begin{equation}
		\frac{n_T(k)}{{(2\pi)^3}} =\frac{N_T}{\zeta(3)} ~\left(\frac{\lambda_{dB}}{2\pi}\right)^3 g_{3/2}\left(\exp\left(-\frac{k^2 \lambda_{dB}^2}{4\pi}\right)\right)
		\label{eqn:th_fun}
	\end{equation}
	wherein the thermal de Broglie wavelength $\lambda_{dB} = \sqrt{2\pi\hbar^2/(m k_B T)}$ yields an estimate of the temperature $T$ which ranges from 100 to 320 nK in our experiments. Here, $g_{3/2}(\cdot)$ is the standard Bose integral, $\zeta(\cdot)$ is the Riemann zeta function, and $N_T$ is the number of atoms in the thermal component. 
	Note that for a non-interacting gas in the thermodynamic limit, the {number of thermal atoms}
	is simply ${N_T^{\rm id}} = \zeta(3)(k_B T / \hbar\bar{\omega})^3{=\eta_T N}$, but for our condensates the critical temperature is reduced by $\approx20\%$ by interactions \cite{PitaevskiiStringari,PethickSmith}. 
	We account for this and the approximately twofold increase in the thermal fraction {$\eta_T$}  (relative to the non-interacting case) by explicitly using $N_T$ as a fit parameter.
    At larger values of momentum, where the thermal component makes a negligible contribution, there appears a slow decay which we identify as the quantum depletion.

\subsection*{Analysis of experimental results}

	A standard approach to analysing the empirical momentum density would be to proceed with a routine fit of the $k$-space histogram with an additional term of the form $C_\alpha/k^\alpha$ to estimate the parameters of the purported quantum-depleted tail.
	If we augment the {thermal} fit function (Eqn. (\ref{eqn:th_fun})) with a power-law term {as per}
	\eq{nkfit}{
	{n(k)} 
	= n_T(k) + \frac{C_{\alpha}}{k^{\alpha}}
	}
	and leave $\alpha$ as a free parameter, the average exponent over all runs is 4.2(4). 
	For comparison, the prior work \cite{Chang16} reported power-law tails with an exponent 4.2(2).
	At first glance, one could simply determine the amplitude of the tails by fixing the exponent to 4, and if we do so, we find an average $C_{\alpha=4}$ which is approximately 8(2) times greater than the coefficient predicted by Eqn. (\ref{eqn:pred_scaling}), and in general agreement with Ref. \cite{Chang16}.
	However, as we detail in the supplementary materials, the covariance of the fit parameters $C$ and $\alpha$, coupled with the exponential relationship to the independent variable $k$, means that this gives 
	a significant underestimate of the uncertainty in $C_\alpha$. 
    In general, fitting  power laws {to data} is known to be prone to return biased estimates of parameters and to drastically under-report uncertainties,\cite{Clauset09,Virkar14} especially when data is available over less than a couple of decades of dynamic range.
	
	Below we 
	present a number of lines of evidence that support the identification of these tails as originating in the quantum depletion, but we also  
	argue there is not sufficient reason to assume that the fit with a fixed $\alpha=4$ is appropriate.
	The main reason for the latter is that the far-field momentum distribution is known to be a modification of the in-situ distribution due to the dispersal of the condensate mean-field energy into kinetic energy. 
	Even neglecting this effect, it is not a given that the far-field distribution could be modified in such a way as to simply increase the amplitude of the tails without otherwise altering the functional form (i.e. the exponent in this case).
	As the authors of Ref. ~\cite{Clauset09} note, ``In practice, we can rarely, if ever, be certain that an observed quantity is drawn from a power-law distribution. The most we can say is that our observations are consistent with the hypothesis that $x$ is drawn from [...] a power law".
	Indeed, this analysis does show that the far-field momentum distribution is consistent with a power-law exponent $3.8\leq\alpha\leq4.6$, but the data at hand cannot precisely determine the exponent $\alpha$ (nor $C_\alpha$), as detailed in the supplement.
		\begin{figure}
	\begin{center}
		\includegraphics[width=\textwidth]{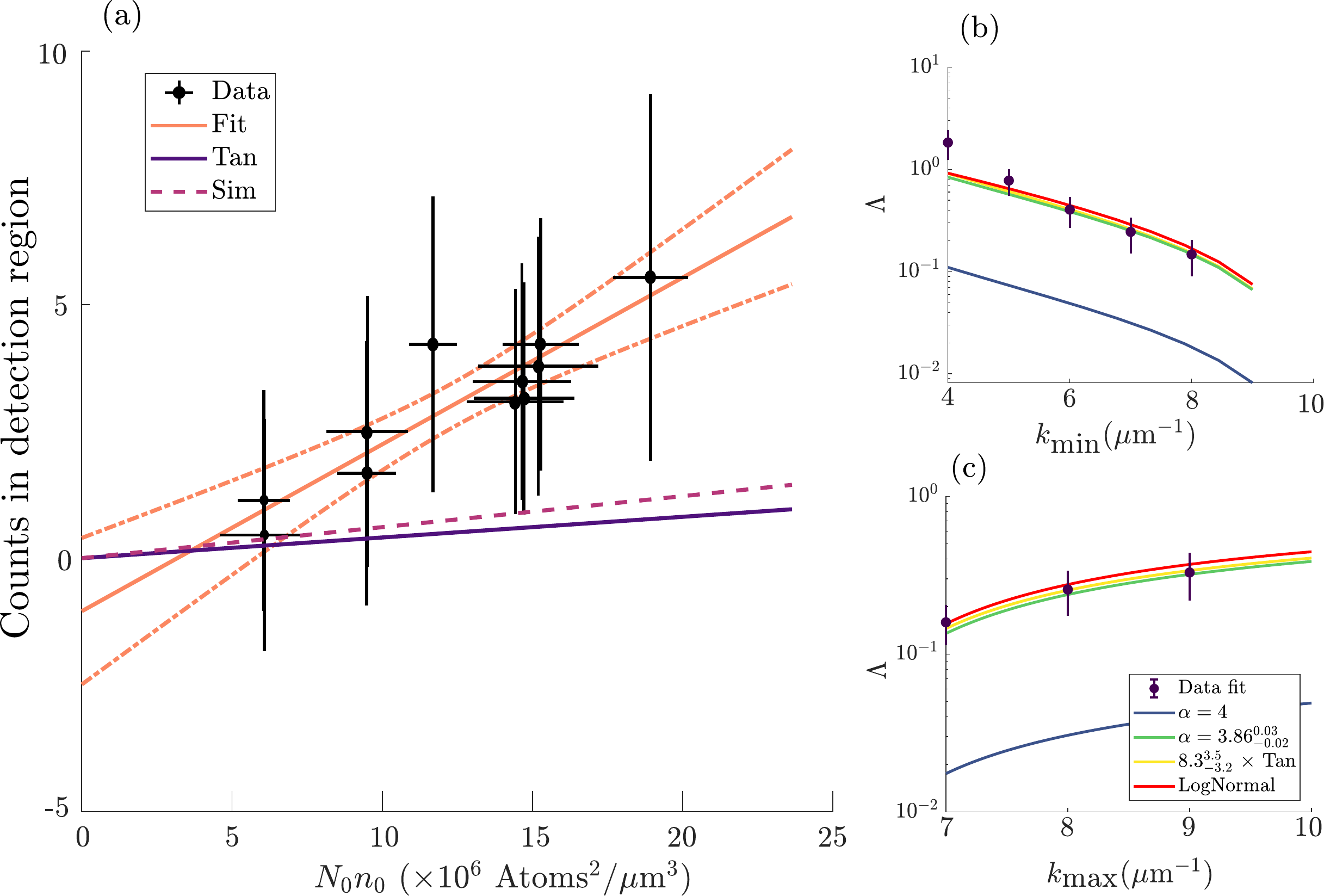}
			\caption{Population of momentum tails, including excess compared to Tan-Bogoliubov theory. (a) The product $N_0n_0$ is a linear predictor of the number of counts within the region $(k_\textrm{min}=6~\micron^{-1},k_\textrm{max}=10~\micron^{-1})$,  consistent with Eqn (\ref{eqn:pred_scaling}) 
			(solid orange line, dashed lines 95\% CI). The gradient {$\Lambda$}
			{in Eqn. (\ref{eqn:Lambda})} can be predicted using Eqn (\ref{eqn:pred_scaling}) ({$\Lambda_{\rm pred}$} solid purple line) but this disagrees with the experiment by a factor of about 8. Our simulations (dashed line, CE in Fig.~\ref{fig:sim_fig}a) show an increase in counts after release but by less than in the experiment. 
			In (b,c) linear fits to the {experimental} data yield {$\Lambda_{\rm fit}$} (points) which vary with the choice of $k$ bounds (fixing $k_\textrm{max}=10\micron^{-1}$ in (b) and $k_\textrm{min}=6\micron^{-1}$ in (c))w.
			For comparison, we
			show predictions of $\Lambda$ based directly on Eqn (\ref{eqn:pred_scaling}) ({$\Lambda_{\rm pred}$}, blue, $n(k)=\mathcal{C}/k^4$), along with the predictions {from Eqn. (\ref{eqn:Lambda}) using} a density function $n(k){=\mathcal{AC}/k^4}$ that {has an additional prefactor $\mathcal{A}=$8(3)} (green) and one that has a modified exponent of {$\alpha=3.86(2)$ via $n(k)=\mathcal{C}/k^{\alpha}$} (yellow). 
			{A log-normal distribution produces nearly identical predictions (red, offset vertically for visibility).}
			{Quoted error estimates correspond to 95\% CI of the fit parameters.}
			In (b), the deviation from the predictions at $k_\textrm{min}\lesssim6~\micron^{-1}$ is because the collection area starts to overlap with the thermal {cloud}.
			}
		\label{fig:exp_results}
	\end{center}
	\end{figure}

	Rather than explicitly enforce a power law decay assumption, 
	we focus on another observable which can be readily measured and predicted: The  number of atoms whose wavevector has a modulus in the interval $k\in (k_\textrm{min}, k_\textrm{max})$,

	\begin{equation}
		N_{k_\textrm{min},k_\textrm{max}} =\frac{\mathcal{C}}{2\pi^2}\left(\frac{1}{k_\textrm{min}}-\frac{1}{k_\textrm{max}}\right)
		\label{eqn:pred_num}
	\end{equation}
	 Note that the integral of $n(k)$ is most easily performed in spherical coordinates and requires the Jacobian $(2\pi)^{-3}{d^3}\kvec$ to ensure normalization. 
	 For fixed $k_\textrm{min}$ and $k_\textrm{max}$, Eqn. (\ref{eqn:pred_num}) has the form 
	 \begin{equation}
	N_{k_\textrm{min},k_\textrm{max}} = \Lambda N_0n_0
			\label{eqn:Lambda}
	\end{equation}
    (c.f. Eqn. {(\ref{eqn:pred_scaling})). We can thus test Eqn. (\ref{eqn:Lambda})} directly by measuring the number of counts detected in the interval $(k_\textrm{min},k_\textrm{max})$ after producing a BEC of $N_0$ atoms with peak density $n_0$. A key advantage of this method is that theoretical assumptions (such as the exponent of the power law) are not required when analysing the experimental data, but only when calculating the (independent) prediction, i.e. the data processing is essentially theory-free.

    Under the null hypothesis (based on the hydrodynamic theory) that the \emph{in situ} depletion does not survive the expansion, $\Lambda=0$.
    {Further, most types of technical noise masquerading as high energy tails would be expected to not follow the $N_0n_0$ scaling and give at best a poor correlation with Eq. (\ref{eqn:Lambda}).} 
	As we show in Fig. \ref{fig:exp_results}, a linear fit of the form $\hat{N}_{k_\textrm{min},k_\textrm{max}} = \Lambda_\textrm{fit} n_0 N_0 + \beta$ yields {an intercept} consistent with zero ($\beta$=-0.9,  95\% CI (-3.1, 1.2)) and a good correlation ($r^2\approx0.8$, $p=1\times10^{-3}$), providing evidence supporting the expected linear relationship with $n_0N_0$, and against the high energy tails being due to some technical noise. 
	The correlation coefficient between the 
	variables {$N_{k_{\rm min},k_{\rm max}}$} 
	and $N_0n_0\propto(N_0^7\bar{\omega}^6)^{1/5}$ is 0.9.
	We conclude that the product $N_0n_0$ is a predictor of the {high energy} 
	population, which is consistent with Eqn. (\ref{eqn:pred_scaling}).
	
	For comparison, a linear fit proves that the atom number {$N_0$} 
	itself is a poor predictor of the detected number ($r^2=0.05~,p=0.54$), as is the {central} density {$n_0$} alone ($r^2=0.4~,p=0.04$).
	Accordingly, the particular nonlinear scaling of detected counts with the predictor $N_0n_0$ is concordant with the tails' originating in the quantum depletion{, and inconsistent with any technical noise that we know of}.
	
	The gradient $\Lambda_\textrm{fit}$ is of particular interest because it can be predicted using Eqn. (\ref{eqn:pred_num}).
	Given
	{a region of interest (ROI) over which we count atoms}, one can calculate $\Lambda_\textrm{pred} = 32 \epsilon a^2(k_{\textrm{min}}^{-1}-k_{\textrm{max}}^{-1})/7$, where $\epsilon$ is the total  detection efficiency. In our experiment, $\epsilon\approx0.23(5)\%$ (see Methods for details).
	In comparison with the predicted value $\Lambda_\mathrm{pred} = 2.7(6)\times10^{-7}$ (units of $\micron^3$/atom), we find that the the empirical fit disagrees with the predicted slope  by a factor of $\mathcal{A}_\mathrm{exp}=\Lambda_\textrm{fit}/\Lambda_\textrm{pred}= 8.3$, 95\% CI $(5.5,11)$, which rules out the null hypothesis.

	While this result may appear to restate the previously-mentioned fitting approach, which gave an increase of the $C_4$ coefficient by a factor of 8(2), it in fact complements it. In this case the overpopulation of the tails is directly measured without any recourse to assumptions about power-law behaviour in the data itself. The direct comparison of the populations in a given $k$-interval allows for an independent comparison between the prediction and measured result and seeks to simply answer the question does the data satisfy the most general model of quantum depletion  proposed by Eq. (\ref{eqn:Lambda}).  
	
	In summary, there are three robust conclusions that can be drawn from the data.
	\emph{First}, the population in the high-momentum tails depends linearly on the product $n_0N_0$, which is a prediction of the Tan and Bogoliubov theories and not readily associated with any other known physical process.
	\emph{Second}, there are some 8(3) times as many particles in the far-field, high-momentum tails as would be expected to be found in the same interval of the in-situ distribution.
	\emph{Third}, the data is consistent with power-laws with exponents in the range $3.8\leq\alpha\leq4.6$.
	
	Interestingly, if one were to take the first observation as sufficient (and indeed independent) evidence to identify the tails with the quantum depletion \emph{and} assume a power-law decay of the form $C_4 k^{-4}$, then one obtains a value of $C_4$ that is consistent with the regression against $n_0N_0$.
	While this is evidence that the $\alpha=4$ hypothesis is not inconsistent with the data, it is essentially the same as calculating $C$ from the results of the linear regression by assuming $\alpha=4$. 
	In Figure \ref{fig:exp_results} (b,c), as a counterpoint to power law fits over $k_{\rm min, max}$ shown in blue, green and yellow, we also show, in red, predictions obtained by assuming log-normally distributed $k$ with parameters $(\mu,\sigma) \approx (1.235,0.95)$ and normalized to the relevant amplitude.
	This underscores the challenge of identifying power-law behaviour in range-limited data, because although the log-normal distribution eventually diverges from the power law, it does so over a much larger domain than available in either Helium experiment (here or \cite{Chang16}).	These fits scarcely differ in their goodness-of-fit criterion (the mean square error) and so offer no obvious way to reconcile the expected distribution with these divergent statistical conclusions.

\subsection*{Findings from simulations}
\label{STAB}

		\begin{figure}[htb]
		\centering
	        \includegraphics[width=\textwidth]{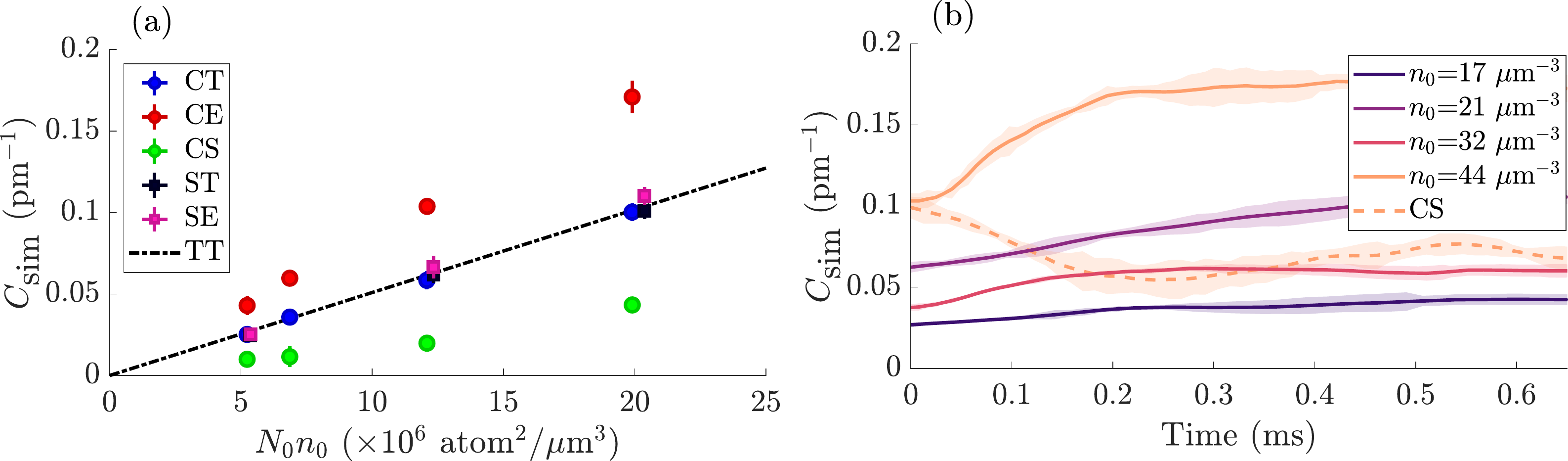}
	        \caption{Simulations of release from the trap. (a) Steady-state values of the simulated contact. Simulations of condensates released from a cigar-shaped trap (CT) are consistent with the Tan theory (TT) before release, and show an increase in contact after the trap release (CE). A slow relaxation of the transverse trapping frequencies (CS) shows a decrease in line with the predicted value of the lower density. Spherical traps (ST,SE) lack any directions of tight confinement, wherein a longer interaction time prevents the escape of depleted particles as seen in cigar traps. (b) the time-dependence of the contact stabilizes after a time on the order of $1/\omega_x$, several hundred $\mu$s. The expanded contact is consistently about 1.7 times the Tan theory. For comparison, the experimental control pulses are implemented after 2ms of expansion. When the transverse trapping frequencies are slowly (1.2ms) reduced by half (dotted line), the in-situ contact relaxes on a faster timescale than the ramp.}
	        \label{fig:sim_fig}
    	\end{figure}

	In order to {understand} whether the depletion could survive the expansion and to investigate what effects are taking place during the initial release, we performed simulations of the BEC expansion from harmonic traps using the first principles STAB method \cite{Deuar11,Kheruntsyan12}. 
	The simulations started from a cigar-shaped trap with parameters matched to the experimental conditions. 
	The in-trap state before release from the trap at time $t=0$ (marked CT in Fig. \ref{fig:sim_fig} (a)) was consistent with the adiabatic sweep theorem applied to the in-situ condensate.
	Following expansion from the cigar trap, the simulated tail amplitude increased and stabilized within a few hundred microseconds (CE in Fig. \ref{fig:sim_fig} (a)), which is much slower than the timescale of the trap potential's vanishing, and implies the far-field tails  stabilize in appearance much sooner than the 2ms delay between the trap release and application of the RF and Stern-Gerlach pulses.
	Fig. \ref{fig:sim_fig}~(b) shows the time evolution of the tail amplitude {$C_{\rm sim}$ extracted from a $n(k)=C_{\rm sim}/k^4$ fit to the simulated density}. 
	In this configuration the steady-state value of the momentum tails was a factor of $C_{\rm sim}/\mathcal{C}=$1.64(9) above the predictions of Eqn. (\ref{eqn:pred_scaling}). 
	{An analysis of the occupation of the tails according to (\ref{eqn:pred_num}), gives very similar factors $\mathcal{A}_{\rm sim}$ for the increase in the strength of the tails (relative to \emph{in-situ} predictions) during evolution, as shown in Supplementary table \ref{tab:Nminmax}.}
	
	To understand the disagreement with earlier theory \cite{Qu16}, which predicted no depletion survival, we also investigated the effect of adiabatic expansion on the in-trap depletion.
    {The characteristic healing timescale $t_{\xi}=\hbar/gn_0=15-40\mu$s in the centre of the trapped cloud is 
    comparable to the trap release time $\tau_{\rm release}$, so a suspicion that adiabaticity is broken in the CE trap release simulations is warranted. For example, $t_{\xi}$  is a characteristic timescale for relaxation of density correlatons due to depletion after a quantum quench \cite{stob}.}
	To test the hypothesis that the difference is due to our system breaking the adiabaticity assumed in \cite{Qu16}, we ran simulations in which the trap is not rapidly released, {but ramped down to half transverse strength} over a much longer time period (CS in Fig.~\ref{fig:sim_fig}). 
	The \emph{in situ} expression (Eqn \ref{eqn:TotalHarmonicContact}) predicts that the depletion should reduce $\propto \bar{\omega}^{6/5}$ to about half its original value. {Indeed it was} found 
	that the in-trap contact {$C_{\rm sim}$ as well as the the tail strength $N_{k_{\rm min},k_{\rm max}}$ from (\ref{eqn:pred_num})} 
	decreased roughly as predicted 
	--- see the dashed line in Fig.~\ref{fig:sim_fig} (b) {and Supplementary table \ref{tab:Nminmax}}, strongly supporting the hypothesis that adiabaticity is needed for agreement with the results of \cite{Qu16}.
	
	{As a check on whether we correctly identify the processes involved in depletion survival, we compared release of atoms from the experimental elongated clouds with spherically trapped clouds having the same central density $n_0$ and particle number $N$. These clouds are labelled (ST,SE) for initial and released clouds, respectively. We find that the survival of depleted atoms is {reduced in the spherical} trap compared to the {elongated ones}. 
	}

\subsection*{{Analysis of simulation results}}

{Our understanding of the above dependencies in the simulations is that the survival and tail strength behaviour are a consequence of the rapid ramp-down of the trap and quench of density which allows escape of non-condensed particles, as well as their acceleration 
}   by the non-uniform mean-field energy of the condensate during the expansion.
   
	In detail, after a quench into the {untrapped} 
	regime, the condensate expands hydrodynamically on timescales of $1/\omega${, and the equilibrium depletion density drops in accordance with falling central density $n_0$ in Eqn. (\ref{eqn:pred_scaling}).} 
	However, whether the actual density in k-space modes follows this equilibrium relationship depends on the reabsorption timescale. Low momentum depletion atoms are unable to escape the condensate before being reabsorbed and are
	absorbed back into the condensate in agreement with \cite{Qu16}.
	However, if reabsorption occurs slower than the change in density, the drop in depletion will be incomplete. High momentum atoms
	have sufficient velocity to escape the expanding cloud 
	without being reabsorbed and thus transition to free atoms. 
	{In our system, as seen in Fig.~\ref{fig:Ck} in the supplement, this concerns particles with wavenumber}  on the order of $k\gtrsim2~\micron^{-1}${, which in particular includes the high momentum tails that are the focus of the experiment. This is the same kind of escape mechanism seen for the appearance of halos of $k, -k$ paired atoms in supersonic BEC collision experiments. \cite{Vogels02b,Krachmalnicoff10,Kheruntsyan12}} 

		Moreover, an atom inside the BEC experiences an effective force from the gradient of the mean-field potential $\textbf{F} = -4\pi\hbar^2 m^{-1}a \nabla  n(x,t)$. 
	This endows escaping depleted particles with a greater momentum.
	{This phenomenon dubbed a ``skiing effect'' \cite{Deuar11b}} has been observed for the thermal part of the cloud in other experiments \cite{Ozeri02,Simsarian00}{, and for supersonic BEC collision halos \cite{Krachmalnicoff10,Deuar14, Hodgman17}.
	For a scale-free distribution such as the $k^{-4}$ power law sought here, such a shift of momentum will manifest itself as an increase of} the amplitude of the tails in the far-field, thus explaining how the observed depletion can appear stronger than in-situ. 
	The simplest very rough estimate of this effect can be made by adding an energy of $gn_0$ to each atom during expansion, obtaining a modified density profile of the form $n(k)\to\approx \mathcal{C}k/(k^2-2gn_0m/\hbar^2)^{5/2}$. This leads, for example, to a doubling of the apparent contact $C_{\rm sim}$ at $k\approx6/\mu$m 
	for clouds with $n_0=39\mu{\rm m}^{-3}$.
    Thus, this modification alone is not sufficient to explain the excess counts in the detection region. 
	
	{A third element is that}
it is much easier for depletion atoms to escape and {the acceleration is larger}  along the tightly-confined axes of a cigar-shaped cloud because the distances $R_{\perp}=(1/\omega_{y,z})\sqrt{2gn_0/m}$ are reduced by $\bar{\omega}/\omega_{y,z}$, whereas the initial mean depletion velocities \textit{in situ} $v\sim \sqrt{2gn_0/m}$ are isotropic.
	Indeed, {the simulations show that} spherical clouds (SE) exhibit a much weaker effect than the elongated clouds (CE) {in agreement with} 
	the longer escape time.
	This {anisotropy effect} also presents as an increase in $C_\textrm{sim}$ {and $\mathcal{A}_{\rm sim}$ for simulation} collection regions (ROI)
	{that include a narrower range of angles around the tight trapping plane.}
	Our ability to test this experimentally was limited because atoms with momenta larger than about 5 $\micron^{-1}$ in the horizontal plane expanded beyond the detector's active surface. 
	Therefore, we obtain only weak evidence of such anisotropy in the experimental data, which is discussed in the supplementary material.

	{The above picture is}
			corroborated by another observation within the simulations:
	During the expansion we observe a decrease in the total number of depleted particles (reabsorption) {as seen in Supplementary Table \ref{tab:simdata} by comparing CE to CT and SE to ST values of $N_B$,} 	and a simultaneous increase of the large-k population (forcing) {described by $C_{\rm sim}$}. 
	A toy model of $k,-k$ depletion modes in a uniform gas undergoing external change in the background density was also investigated to verify our interpretation of the processes involved.

\subsection*{Toy model of escape}

\begin{figure}[htb]
		\centering
	        \begin{center}
		\includegraphics[width=0.3\textwidth]{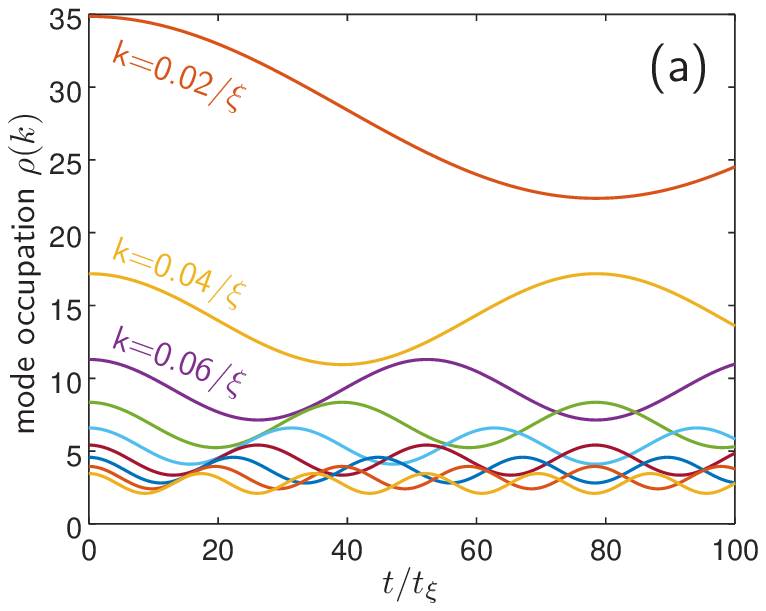}
		\includegraphics[width=0.3\textwidth]{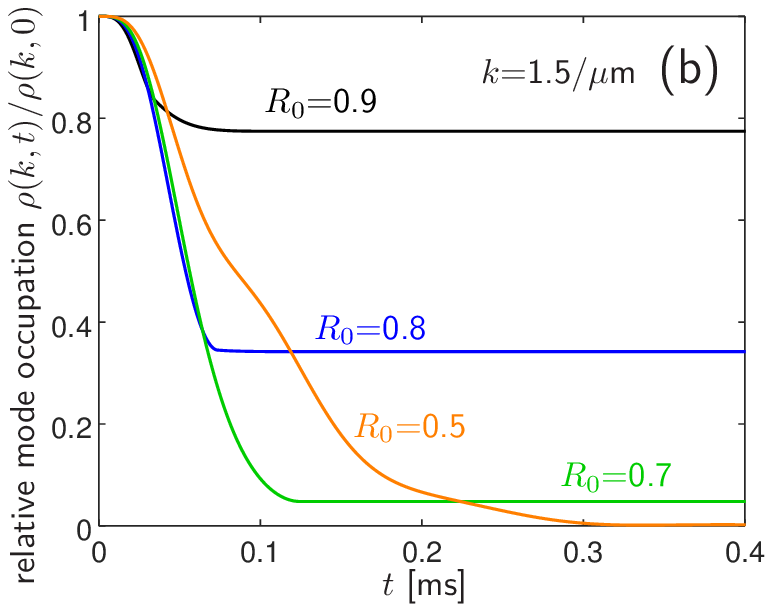}\\
		\includegraphics[width=0.3\textwidth]{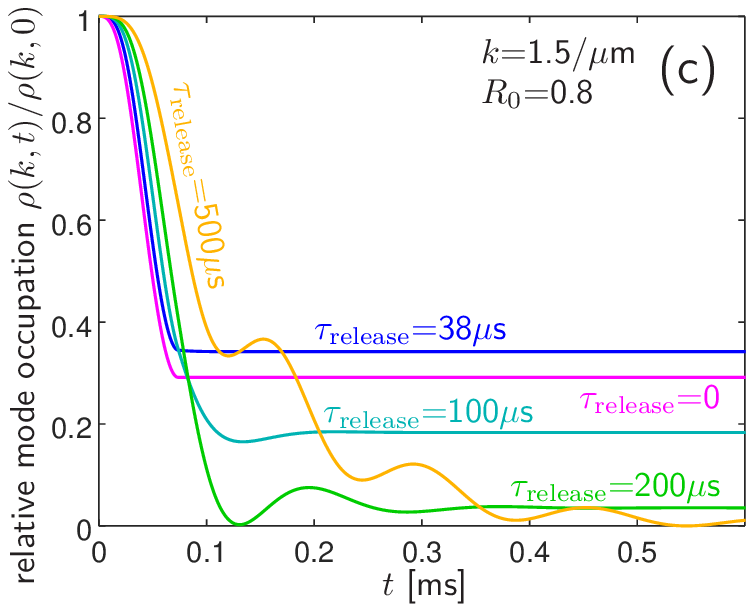}
		\includegraphics[width=0.3\textwidth]{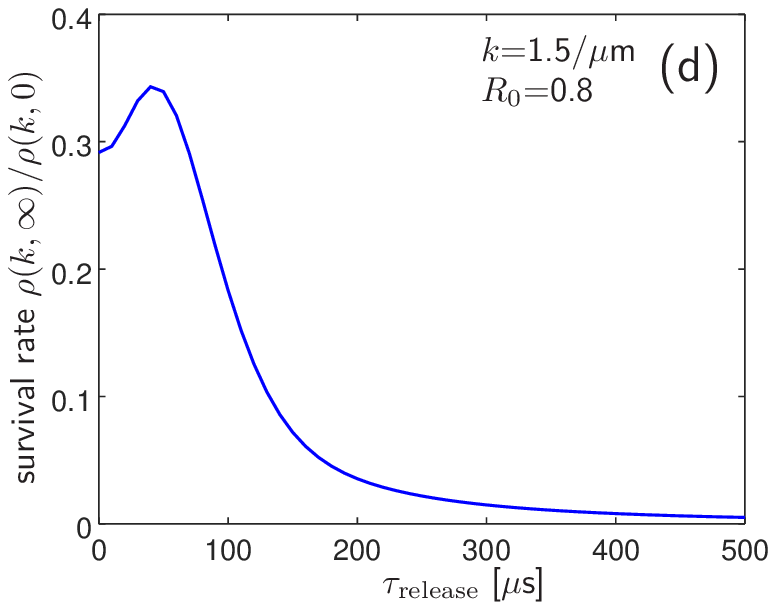}
		\includegraphics[width=0.3\textwidth]{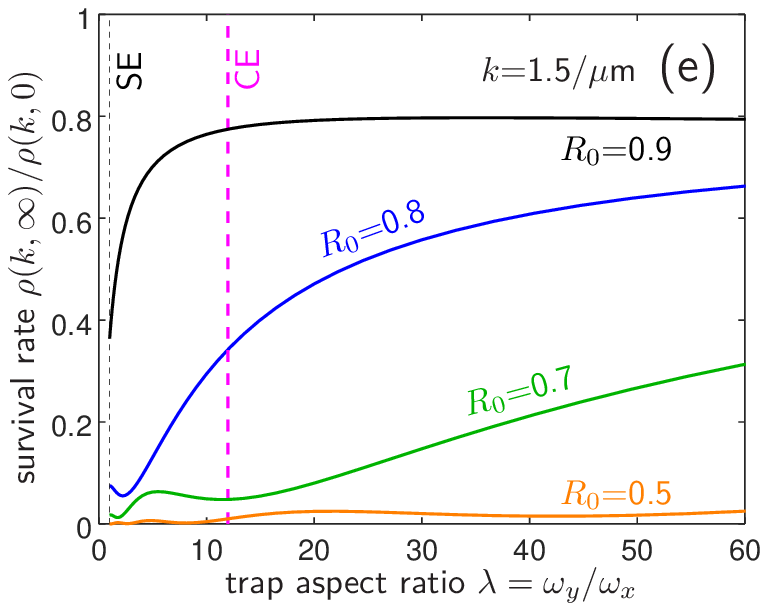}
		\end{center}\vspace*{-0.5cm}
	        \caption{Toy model simulations for $k$ and $-k$ modes in a uniform gas initially in the Bogoliubov ground state. 
	        Panel (a): behaviour after a ``caricature'' quench to half-density, as per Eqn. \eqn{eq:caricature} for modes with $k\xi=0.02,0.04,\dots,0.18$. 
	        The remaining panels concern the better toy model of Eqs. (\ref{eq:toy},\ref{toy:nt}) with parameters like the full simulation with $\omega=902\times895\times71$ Hz, $N=455852$, $n_0=43.66/\mu m^3$ peak density, and $\tau_{\rm release}=38\mu$s, and show the mode occupation evolution relative to the initial value (the survival rate). 
	        Panel (b): For different initial locations in the condensate in the narrow direction: $R_0=y/R_{\perp}$ where $R_{\perp}=(1/\omega_y)\sqrt{2gn_0/m}$, and initial $k=1.5/\mu$m. 
	        Panel (c): 
	        Evolution of the 
	        relative occupation for different ramp speeds $\tau_{\rm release}$. Cycles of reabsorption are seen for the slow ramps.
	        Panel (d): 
	        final survival rate for the same parameters.
	        Panel (e): 
	        shows the dependence of the final survival rate 
	        on the trap aspect ratio $\lambda$, when initial central density $n_0$ is kept constant. The magenta dashed line indicates the experimental $\lambda=12$ (like CE simulations), the black dashed line a spherical trap $\lambda=1$ (like SE simulations). 
	        }
	        \label{fig:toymodel}
    	\end{figure}

	The reabsorption mechanism and qualitative features of the escape of depletion from the condensate discussed above can be seen in a toy model of two $k$ and $-k$ Bogoliubov modes in a uniform volume of gas at zero temperature when the background density is quenched due to external factors, as described in the supplementary material. 
	The simplest such ``caricature'', when the density $n_0$ is quenched to $n'<n_0$ at $t=0$ and then stays constant, has the occupation of each $k$ mode evolve as
	\begin{equation}
	\rho(k,t) = \rho(k,0) - \frac{gn[\epsilon_0(k)^2-\epsilon(k)^2]}{4\epsilon(k)^2\epsilon_0(k)}\,[1-\cos 2\epsilon(k) t].
	    \label{eq:caricature}
	\end{equation}
	Here, $\epsilon_0(k)$ and $\epsilon(k)$ are given by \eqn{bogfreq} using the initial $n_0$ and later $n$ values of density, respectively. 
	The reabsorption comes about then via the initial dip of the Rabi oscillations seen in Fig.~\ref{fig:toymodel}(a). The Rabi oscillations are between the two superpositions corresponding to the initial Bogoliubov ground state and the final one.
	However, in the actual expansion it can happen that later (return) stages of the Rabi oscillations never eventuate if the density drops faster than the oscillation frequency; roughly $\omega_{\perp}\gtrsim2\epsilon(k)$.

	Figs.~\ref{fig:toymodel}(b-e) 
	show the behaviour of a more careful toy model given by Eqn. \eqn{eq:toy} and \eqn{toy:nt} (requiring numerical integration) in which the background density decays in a time-dependent fashion that at least qualitatively approximates a significant part of what happens during release. 
	Panel (b) 
	concerns escape of particles that 
	start in the outer parts of the cloud  ($R_0=y/R_{\perp}\gtrsim0.5)$. 
    Panels (c-d) 
    show the dependence of survival on the speed of the ramp which turns off the trap, indicating that ramp speeds $\tau_{\rm release}\lesssim100\mu$s are mostly neutral for the effect but slower ramps strongly suppress the escape. 
    Panel (e)
    considers the survival rate for different trap aspect ratios, including the case of $\lambda=12$ like in the experiment (CE) and the spherical case (SE). Survival is greatly aided by elongated traps.
	
	There are still many effects missing from the toy model compared to the STAB simulations (high momentum atoms on trapped trajectories temporarily outside the condensate at time of release, multidirectional flight of the depleted atoms, reduced skiing due to simultaneous collapse of the condensate density, energy-momentum uncertainty, the effect of the remnant trapping potential on depletion atoms, to name a few) and survival rate is lower than in the full 3D simulations. It does, though, give a first qualitative underpinning for the escape effects seen in the full many-mode simulations.

\section*{Discussion}

\label{sec:discussion}

	We find that the number of atoms in the large-$k$ tails in the far-field is consistent with the tail amplitude scaling as a linear function of the product $N_0n_0\propto(N_{0}^7\bar{\omega}^6)^{1/5}$, in line with Tan's theory of the contact (Eqns. \ref{eqn:pred_scaling},\ref{eqn:pred_num}). 
{Neither thermal populations nor a number of technical effects (imaging, single particle processes, background noise) can be expected to have the same $N_0n_0$ scaling.}
	However, the effect size is significantly different than naively expected from in-situ values by a factor of order 8(3), and from simulated values by a factor of 5(3) which is not accounted for by any 
	obvious systematic effects.
	
	The earlier experiment by Chang \emph{et al.} \cite{Chang16} also noted the $N_0n_0$ scaling and an excess by a factor of about 6 that falls within  our error bars. However, recent investigations \cite{Cayla22} found that this was correlated with the presence of a small impurity fraction consisting of $m_{j}=0$ atoms ($\sim1$\%) in their optically trapped cloud, which was otherwise spin polarised in the $m_{j}=1$ state. When the impurity fraction was reduced to $0.05$\% tail survival was no longer observed. In contrast, our experiment is conducted in a magnetic trap, which is unable to confine any spin state other than the $m_{j}=1$ state.  Thus our results can not be explained by the presence of similiar trapped impurities.

{While} the thermal quasiparticles in the Bogoliubov picture simply map onto the thermal population of constituent particles of the same momentum (See, for example, \cite{PethickSmith} Chap. 8.3. or \cite{Vogels02}), 
	{a remnant	thermal population is not a good candidate to explain the observations. This is because it} decays super-exponentially with $k$ {(Bose-Einstein distribution)}, and hence does not account for the atoms we observe beyond $k\gtrsim 6~\micron^{-1}$, even though it is subject to the same mean-field forcing as the depletion \cite{Ozeri02}.
	We can show this with a simple calculation, noting that on physical grounds the maximum energy that can be imparted {by the ``skiing effect''}\cite{Deuar11b} is $\mu=gn_0$, where $n_0$ is the initial density in the centre of the cloud.
	For an atom with momentum $k=6\micron^{-1}$, at the edge of the thermal region in the densest cloud we consider (44$\micron^{-3}$), the additional energy $\mu$ imparts at most a momentum shift of order 0.7$\micron^{-1}$, which is insufficient to account for the population detected as far out as $k=10\micron^{-1}$.
The phonon/particle changeover is {also} not responsible for the inflections seen at high $k$ in Figure~\ref{fig:nk_density} 
because this changeover occurs at $k\sim1/\xi\approx 2\mu{\rm m}^{-1}$.
	
   Another candidate explanation to consider would be large depletion produced after release in the short-lived mixed-species condensate (where the $m_J=0,1$, and -1 clouds overlap after the Landau-Zener sweep). 
   {This could have the observed $N_0n_0$ scaling.}
	{However,} the expression for the contact in a mixed-species bosonic gas \cite{Werner12_boson} can be combined with the energy of a condensed mixture \cite{PethickSmith} to show that the contact in spin-mixed systems is bounded from above by the contact of  the same system polarized in the most strongly-interacting state \cite{Braaten11}.
	The inter-spin scattering lengths $a_{ij}$ (between \mhe atoms in the $i$ and $j$ spin states) are not fully characterized by experiments, but can be estimated to be $a_{11}=a_{-1-1}=a_{01}=a_{0-1}\approx 140~a_0$, $a_{00}=120~a_0$ and $a_{1-1}\approx60~a_0$, in terms of the Bohr radius $a_0$\cite{Vassen16}. 
	Thus, the contact in a \mhe condensate is maximized when the cloud is purely polarized in the $m_J=1$ state.
	Any mixture of {He${}^*$} spin states is thus predicted to have a lower contact (and thus less-populated tails) than the initial condensate, {which appears to rule out this route to explain} 
	our observations.

On the other hand the simulations and toy model
	demonstrate a route for escape of the fast depletion atoms from the cloud, and indicate 
	that the survival of the quantum depletion into the far-field is possible when the release is non adiabatic, 
	but not as a straightforward mapping into the far-field density. The latter is due to 
which imparts some acceleration to the atoms during the early stages of the expansion \cite{Ozeri02}.

We thus {are led to} surmise that the experimentally observed tails are indeed a remnant of the 
 quantum depletion (per the {observed scaling with $N_0n_0$ and $\approx k^{-4}$ that matches the Tan theory, qualitative similarity in behaviour to the simulations, and lack of convincing counter-hypotheses}),
albeit subject to some physical effect during the expansion or some nonequilibrium enhancement in the trapped state.

	In conclusion, 
	we find statistically robust evidence that the quantum depletion can, remarkably, survive the expansion and dilution of its original condensate under certain conditions. 
	Our simulations also demonstrate a mechanism by which the uncondensed quantum depleted atoms of a single species 
	can be visible in the far-field momentum distribution, 
	and that the hydrodynamic approximation does not capture sufficient short-wavelength information to make detailed predictions about the high-momentum behaviour. 
	We thus find a partial explanation for the experimental deviation of the far-field distribution from both the in-situ and the hydrodynamic pictures, although there is an unexplained discrepancy at this time between theory and experiment as to the amount of this growth. The results reported here expand the growing body of data and knowledge regarding the somewhat mysterious behaviour of the far-field quantum depletion \cite{Cayla20,Chang16,Qu16,Cayla22}.

Since the exact mechanism responsible for the impurity effects seen in \cite{Cayla22} remains unrecognised, it is uncertain whether it can also be responsible for our measurements of tail strength well in excess of the single-species simulation. Our experiment does not involve impurities in the initial trapped cloud.

\section*{Methods}

\subsection*{Experimental setup} 

\begin{figure}[htb]
	\centering
	    \includegraphics[width=0.5\textwidth]{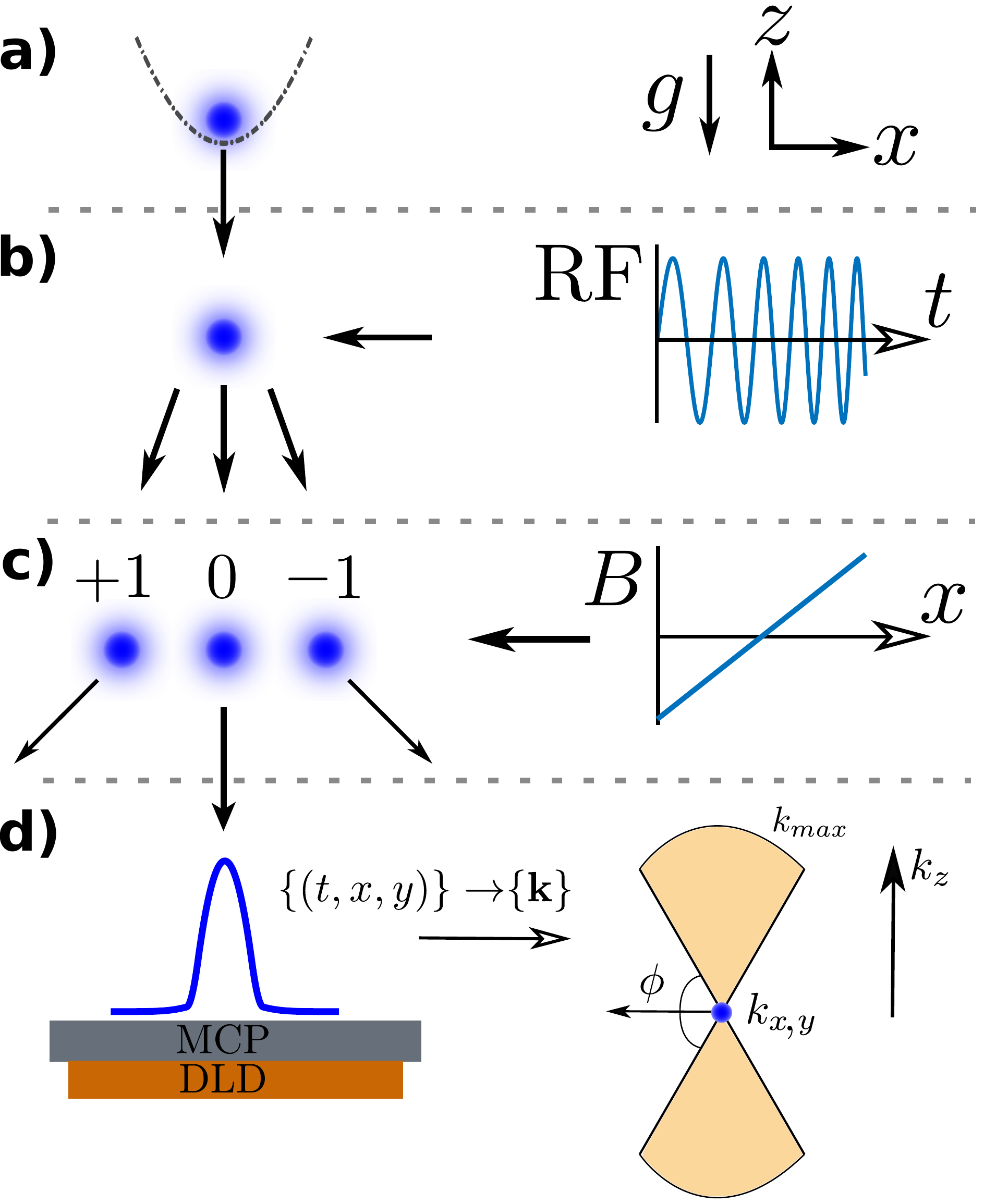}
	    \caption{Sketch of the experimental sequence. A BEC is released from a harmonic trap 
	    (a) and expands during freefall before being split into a superposition of the $m_J\in\{-1,0,1\}$ states (b) by an RF chirp. A magnetic field gradient separates the clouds (c) ensuring that only the magnetically insentitive $m_J=0$ cloud lands on the detector (d), from which the momentum information is reconstructed. Due to the finite detector radius, the collection region in momentum-space is restricted to two vertically-oriented spherical segments (shaded region) whose boundary subtends an angle ${\phi_c}=\pi/3$ with the horizontal $(x,y)$ plane. The quantum depletion lies in
	    the dilute tails at large momentum $\gtrsim 6\micron^{-1}$ (see Fig. \ref{fig:nk_density})}
	    \label{fig:sequence}
	\end{figure}

    Our experimental sequence {for measurement runs as described in the ``Experimental meaurements'' section above is shown schematically in Fig.~\ref{fig:sequence}.} 

	We prepared our BECs via forced evaporative cooling in a harmonic magnetic trap with trap frequencies $\approx(45,425,425)$ Hz and a DC bias stabilized by our auxiliary field compensation coils \cite{Dall07,Dedman07}. For the tight trap we increased the coil current after the cooling sequence to obtain trapping frequencies $\approx(71,902,895)$ Hz, ramping the current as a sigmoid step function to minimize in-trap oscillations. Note that the weak ($x$) axis of the trap is horizontal, with tight vertical confinement. 
	The RF pulse was created by a  function generator, amplified, and applied to the experiment chamber by a coiled antenna inserted into the BiQUIC coil housing. The pulse swept from 1.6-2.6MHz over 1ms and was centred on the resonance between the $m_J$ states. The determination of the transfer efficiencies $\eta_J$ for each of the $m_J$ states is discussed below. The sweep was $10^6$-fold wider than the RF width of the BEC to ensure uniform transfer at all momenta. Immediately after the RF sweep, the bias coils are switched off and auxiliary push coils in the vertical (Z) and weak horizontal (X) axes are activated using a fast MOSFET switch to implement a Stern-Gerlach deflection of the $m_J = -1,$ and $+1$ atoms, such that only $m_J=0$ state atoms reach the detector.
		The Stern-Gerlach (SG) pulse was designed by increasing the pulse duration until the $m_j=\pm 1$ clouds were given sufficient velocity to reach the edges of the detector ($\approx 10$ cm/s), and then doubling the current passed through the field-generating coils.

	We use an 80mm diameter multichannel plate and delay-line detector stack \cite{Manning10} located 848mm below the trap, which registers the arrival times and positions $(t,x,y)$ of each atom. 
	The velocity of each atom relative to the centre of mass of each cloud is calculated by $(v_x,v_y,v_z) = t_{i}^{-1}(x_i-\bar{x},y_i-\bar{y},\tfrac{1}{2}g_0(t_{cen}^2-t_{i}^{2}))$, where $g_0$ is the local gravitational acceleration, the overbar denotes the within-shot average and $t_{cen}$ is the time of flight of the centre of mass of the cloud. 
	The far-field momentum is thus obtained via $m\textbf{v} = \hbar\kvec$, noting that this cannot be identified with the in-situ momentum (see Discussion section).
	The space and time resolution of the detector are 100 $\mu$m and 3 $\mu$s, respectively \cite{Henson18}.
			Sets of ten experimental runs were interleaved with calibration measurements to determine the shot-to-shot variation in atom number, trapping frequencies, magnetic state transfer efficiency, and noise contributions in the manner described in the supplementary material.

	The detector {quantum} efficiency {(QE)} of ${\eta_Q=}8(2)\%$ was determined from analysis of the squeezing parameter of correlated atoms on the opposite sides of scattering halos \cite{shin19,shin20,Jaskula10}. 
	{A second factor affecting the total collection efficiency $\epsilon$ is that}
	the $k$-space field of view is restricted by the detector radius to $k\lesssim5 /\mu{\rm m}$ in the $(x,y)$ plane, which is only just sufficient to reach past the edge of the thermal region. 
	We thus face a tradeoff in the choice of $k_\textrm{max}$, therefore we define the bounds of our region of interest (ROI) by the minimum elevation angle $\phi_c=\pi/3$ rad above the $(x,y)$ plane and an upper bound of $k_\textrm{max} = 10\micron^{-1}$ {(beyond which the signal to noise ratio becomes too poor)}. This amounts to an ROI consisting of two vertically oriented spherical segments, each with half-angle $\pi/6$ from the $z$ axis, encompassing a total solid angle of ${\Omega_{ROI}=4\pi(1-\sin{\phi_c})=}0.13\times 4\pi$ steradians. 
	
	We must also account for the state-transfer efficiency of ${\eta_0=}25(2)$\% {during the RF sweep}, and combine all these factors into the total efficiency $\epsilon{=\eta_Q\eta_0(1-\sin{\phi_c})}\approx0.23(5)\%$.
{
	{The dominant uncertainty in the collection efficiency $\epsilon$ is the 25\% error in the detector quantum efficiency (QE), whereas the other factors (cutoff angle $\phi_c$ and transfer efficiency $\eta_0$) are more precisely known.
	
	We performed the analysis {of the depletion tails} described above for a range of $\phi_c$ and values of the QE {$\eta$} and found that the excess of counts (expressed as $\Lambda_\textrm{fit}/\Lambda_\textrm{pred}$) was not significantly affected. {This is summarized} Supplementary Table \ref{tab:choice_indep}.}}

\subsection*{{Simulations}} 

	The STAB method (stochastic Time-Adaptive Bogoliubov) \cite{Deuar11,Kheruntsyan12} uses the positive-P representation \cite{Drummond80,Deuar07} to describe Bogoliubov quasiparticles around a dynamically evolving condensate \cite{Sinatra00}. This allows for straightforward treatment of inhomogenous and evolving condensates with their associated quantum depletion, without the need to diagonalise the Bogoliubov-de Gennes equations. The systems considered here require $4-6\times10^6$ modes for simulation, so avoiding diagonalisation is very relevant.	Previous use of the STAB method\cite{Kheruntsyan12,Deuar11,Krachmalnicoff10,Jaskula10,Lewis-Swan14,Lewis-Swan15,Deuar13,Deuar14} has been according to the equations described in detail in \cite{Deuar11} which relied on a separation of the condensate and Bogoliubov quasiparticles in k-space that arose from initial conditions and system dynamics. Here this does not occur, and there is a significant overlap in momentum space. The standard STAB formulation leads to an unphysical amplification of the part of the Bogoliubov field that overlaps with the condensate. Therefore a theory that explicitly imposes orthogonality between condensate and Bogoliubov modes is required. We summarise our approach {here, with some technical details} in the supplement{al material.} 
	Details of the derivation and proper benchmarking of the {modified} method will be reported in \cite{nstab-longpaper}.
	
\subsubsection*{{Orthogonalised STAB method}}

	In terms of operators, the Bose field of the atoms $\op{\Psi}(\bo{x},t)$ is written as
	\eq{bog}{
	\op{\Psi}(\bo{x},t) = \phi(\bo{x},t) + \op{\Psi}_B(\bo{x},t),
	}
	where $\phi(\bo{x},t)$ is the condensate order parameter described in 3-dimensional space $\bo{x}$, and $\op{\Psi}_B(\bo{x},t)$ is a relatively small operator fluctuation field. 
	The smallness requirement can be written 
	\begin{align}
		N=&\int d^3\bo{x}\ |\phi(\bo{x},t)|^2 \\
		\gg& \int d^3\bo{x}\left\langle\dagop{\Psi}_B(\bo{x},t)\op{\Psi}_B(\bo{x},t)\right\rangle=N_B=N\delta_B,
		\label{eqn:exp}
	\end{align}
	i.e. $N_B$ the number of particles in the Bogoliubov field is small overall, but locally the Bogoliubov field  density need not be smaller then the condensate {-- $\delta_B$ is the small parameter of the theory \cite{Castin98}}. 
	The condition \eqn{eqn:exp} allows one to discard third and higher orders of $\op{\Psi}_B$ in the effective Hamiltonian (the Bogoliubov approximation).
	A second condition, not applied in standard STAB, {but present in more precise flavours of Bogoliubov theory} is
	\eq{orth}{
	\int d^3\bo{x}\ \dagop{\Psi}_B(\bo{x},t)\phi(\bo{x},t)= 0.
	}
	which  imposes orthogonality and prevents seeping of condensate atoms into the fluctuation field $\op{\Psi}_B(\bo{x},t)$.

	The condensate order parameter $\phi(\bo{x},t)$ is assumed to evolve according to the Gross-Pitaevskii equation (correct to leading order, given \eqn{eqn:exp}):
	\eqa{GPE-eq}{
	i\hbar\frac{d\phi}{dt} = \left[-\frac{\hbar^2}{2m}\nabla^2+g|\phi|^2+V(\bo{x},t)\right]\phi.
	}
	and is normalised to the (conserved) total number of particles $\int d^3\bo{x}\ |\phi(\bo{x},t)|^3=N$. The $g=4\pi\hbar^2a_{1,1}/m$ is the s-wave contact interaction between He${}^*$ atoms in the initial $m_J=1$ state (we take $a_{1,1}=7.51$nm), and $V(\bo{x},t)$ is the trap potential with in general time-dependent frequency. 
	We then represent the Bogoliubov quasiparticles using the positive-P representation \cite{Drummond80,Deuar11}, which leads to the following
	equations of motion: 	

	\leqs{pstab-eq}{
	i\hbar\frac{d\psi_B}{dt} &=&  \left[-\frac{\hbar^2}{2m}\nabla^2+g|\phi|^2+V(\bo{x},t)\right]\psi_B
	+\mc{P}_{\perp}\left\{
	g|\phi|^2\psi_B + g\phi^2\wt{\psi}_B^* + \sqrt{-ig}\ \phi\,\xi(\bo{x},t)
	\right\}\\
	i\hbar\frac{d\wt{\psi}_B}{dt} &=& \left[-\frac{\hbar^2}{2m}\nabla^2+g|\phi|^2+V(\bo{x},t)\right]\wt{\psi}_B
	+\mc{P}_{\perp}\left\{
	g|\phi|^2\wt{\psi}_B + g\phi^2\psi_B^* + \sqrt{-ig}\ \phi\,\wt{\xi}(\bo{x},t)
	\right\}.
	}
	Here the ket $\psi_B(\bo{x},t)$ and bra $\wt{\psi}_B(\bo{x},t)$ amplitudes provide the positive-P representation of the Bogoliubov field {$\hat{\Psi}_B(\bo{x},t)$} in 3D space. We used the robust stochastic integration procedure described in \cite{tcorr}. The $\xi(\bo{x},t)$ and $\wt{\xi}(\bo{x},t)$ are independent white Gaussian noise fields of zero mean and variance:
	\eq{noises}{
	\langle\xi(\bo{x},t)\xi(\bo{x}',t')\rangle = \langle\wt{\xi}(\bo{x},t)\wt{\xi}(\bo{x}',t')\rangle = \delta^3(\bo{x}-\bo{x}')\delta(t-t').
	}
	An ensemble of field trajectories with independent noise in each trajectory and in each trajectory's initial state is generated to represent the Bogoliubov field. We typically used $\mc{S}=4000$ trajectories. 
	Notably, the equations \eqn{pstab-eq} allow not only for production of additional Bogoliubov quasiparticles quantum depleted from the condensate but also for their reabsorption.
	The main additional element in \eqn{pstab-eq} compared to the standard STAB equations \cite{Deuar13} is the projection $\mc{P}_{\perp}$ which imposes the orthogonality requirement \eqn{orth} and avoids the aforementioned amplification of the Bogoliubov field where it overlaps with the condensate. 
	The projection $\mc{P}_{\perp}$ of a field $f(\bo{x})$ can be carried out efficiently by 
	\eq{Pc}{
	\mc{P}_{\perp} f(\bo{x}) = f(\bo{x})- \frac{1}{N}\left[\int d^3{\bo{x}'}\ \phi({\bo{x}'})^*f({\bo{x}'})\right]\,\phi(\bo{x}).
	}
	The kinetic part of the evolution equations \eqn{GPE-eq}-\eqn{pstab-eq} is also carried out efficiently by a split-step approach which evaluates kinetic terms in k-space and the rest in x-space, moving between k-space and x-space using a fast Fourier transform.
	{Calculation of observables is described in the supplementary material.}

	\subsubsection*{Initial condition}
    \label{TH-INI}
	Our simulations aim to study the evolution of the quantum depletion particles in $\op{\Psi}_B$ after release from the trap. 
	We use a zero temperature initial condition, since the object is to study the behaviour of the high momentum tails beyond the edge of the thermal cloud{, in which $T>0$ effects are negligible.} 
	The $T=0$ initial state is {also} more straightforward to obtain, {allowing one} 
	to use lower k-values to access the $k^{-4}$ tails, since they are not obscured by the stronger thermal cloud at intermediate momenta. 
	This significantly reduces the size of the computational lattice needed.
	For the low temperatures in the experiment we do not expect any significant interaction between the behavior of the thermal cloud and the depleted atoms because both are well approximated by the Bogoliubov {approximation} 	which neglects interactions between excited modes. Therefore the neglect of the thermal cloud does not significantly affect the properties of the higher k depletion or its evolution.

	However, one cannot use the standard Gross-Pitaevskii ground state since that has 100\% condensate and no quantum depletion. 
	The task of generating a cloud with the appropriate depletion in such a large nonuniform system turns out to be nontrivial. 
	Conceptually the issue is simple -- diagonalise the Bogoliubov Hamiltonian, and give the well known Bogoliubov $T=0$  occupation to each quasiparticle mode. 
	However, for a system with $10^6$ modes diagonalisation is not a good option.
	Our solution to this situation is 
	{to make a calibrated quantum quench from the Gross-Pitaevskii solution to the full Bogoliubov equations of motion, which provides a state with an appropriate quantity of quantum depletion. The technique is} described in detail in the supplementary material.

	\subsubsection*{Simulation types}

\noindent 	Several types of simulations were made, with shorthand labels as per Fig. \ref{fig:sim_fig}{, and summarised in Supplementary Table \ref{tab:simdata}:}
	
	\textbf{(CE)} Release of atoms from the trap, as in the experiment. Here the potential was reduced exponentially
	\eq{Vt}{
	V(\bo{x}) = \frac{m}{2}\left(\omega_x^2x^2+\omega_y^2y^2+\omega_z^2z^2\right)\,e^{-t/\tau_{\rm release}},
	}
	with time constant $\tau_{\rm release}=37.5\mu$s, matched to the experiment. The initial trap frequencies were  $\omega=425\times425\times45$ Hz and $\omega=902\times895\times71$ Hz, and two variants of the initial state were simulated: a low density and a high density cloud.

	\textbf{(CS)} {Slow decrease of the transverse trapping frequencies by a factor of two. Here we}
	ramped the trap as follows:
	\eq{Vtramp}{
	V(\bo{x}) = \frac{m}{2}\left[\left(\omega_x^2x^2+\omega_y^2y^2\right)\left(1-\frac{t}{2t_{\rm ramp}}\right)^2 +\omega_z^2z^2\right],
	}
	with timescales of order 1-2ms (see Supplementary Table \ref{tab:simdata}). The simulations were run up till $t=t_{\rm ramp}$ when the {transverse} trap frequency was half the original one.

	\textbf{(ST,SE)} {Trap release of spherical clouds.} 
	The velocity distribution of the depletion atoms is isotropic in situ, being given by $mv^2/2 \approx gn_0$. 
	However, the distance to travel to escape reabsorption depends on the cloud shape. In particular, escape is made easier in the tight trap directions (less distance to travel), and harder in the long trap direction. 
	{Here we used} spherically trapped clouds having the same central density $n_0$ and particle number $N$. These clouds had isotropic trapping frequency $\wb{\omega}=(\omega_x\omega_y\omega_z)^{1/3}$ and are labelled (ST). {Trap release (SE)} followed \eqn{Vt} as before.

\section*{Acknowledgements} 
We Would like to thank David Clement, Jean Dalibard, Karen Kherunstyan, and Raphael Lopes for their helpful discussions. This work was supported by Australian Research Council (ARC) Discovery Project Grants No. DP160102337 and No. DP190103021. S. S. H was supported by DECRA DE150100315,  J.A.R., D. K. S. by the Australian Postgraduate Award (APA), and K.F.T. by the Australian Government Research Training Program (RTP) Scholarship. The simulations by P. D. were supported by National Science Centre (Poland) grants No. 2018/31/B/ST2/01871 and 2012/07/E/ST2/01389.

\section*{Author contributions statement}

The data was collected and analysed by J. A. Ross, D. K. Shin, K. F. Thomas, and B. M. Henson under the supervision of S. S. Hodgman and A. G. Truscott. P. Deuar conducted the simulations. All authors contributed to the interpretation of results. J. A. Ross and P. Deuar wrote the paper with input from all authors.

\section*{Additional information}

The authors declare no competing interests. The data generated and analysed during the current study are available from the corresponding author on reasonable request.

\bibliography{main}

\begin{thebibliography}{10}
\urlstyle{rm}
\expandafter\ifx\csname url\endcsname\relax
  \def\url#1{\texttt{#1}}\fi
\expandafter\ifx\csname urlprefix\endcsname\relax\def\urlprefix{URL }\fi
\expandafter\ifx\csname doiprefix\endcsname\relax\def\doiprefix{DOI: }\fi
\providecommand{\bibinfo}[2]{#2}
\providecommand{\eprint}[2][]{\url{#2}}

\bibitem{Bogolubov47}
\bibinfo{author}{Bogoliubov, N.}
\newblock \bibinfo{journal}{\bibinfo{title}{On the theory of superfluidity}}.
\newblock {\emph{\JournalTitle{Journal of Physics USSR}}}
  \textbf{\bibinfo{volume}{XI}}, \bibinfo{pages}{23} (\bibinfo{year}{1947}).

\bibitem{Vogels02}
\bibinfo{author}{Vogels, J.~M.}, \bibinfo{author}{Xu, K.},
  \bibinfo{author}{Raman, C.}, \bibinfo{author}{Abo-Shaeer, J.~R.} \&
  \bibinfo{author}{Ketterle, W.}
\newblock \bibinfo{journal}{\bibinfo{title}{Experimental observation of the
  {Bogoliubov} transformation for a {Bose-Einstein} condensed gas}}.
\newblock {\emph{\JournalTitle{Phys. Rev. Lett.}}}
  \textbf{\bibinfo{volume}{88}}, \bibinfo{pages}{060402},
  \doiprefix\url{10.1103/PhysRevLett.88.060402} (\bibinfo{year}{2002}).

\bibitem{PitaevskiiStringari}
\bibinfo{author}{Pitaevski\v{i}, L.~P.} \& \bibinfo{author}{Stringari, S.}
\newblock \emph{\bibinfo{title}{{Bose-Einstein} condensation and
  superfluidity}}.
\newblock No. \bibinfo{number}{164} in \bibinfo{series}{International series of
  monographs on physics} (\bibinfo{publisher}{Oxford University Press},
  \bibinfo{address}{Oxford, United Kingdom}, \bibinfo{year}{2016}),
  \bibinfo{edition}{first edition} edn.
\newblock \bibinfo{note}{OCLC: ocn919186901}.

\bibitem{Decamp18}
\bibinfo{author}{Decamp, J.}, \bibinfo{author}{Albert, M.} \&
  \bibinfo{author}{Vignolo, P.}
\newblock \bibinfo{journal}{\bibinfo{title}{{Tan's} contact in a cigar-shaped
  dilute {Bose} gas}}.
\newblock {\emph{\JournalTitle{Phys. Rev. A}}} \textbf{\bibinfo{volume}{97}},
  \bibinfo{pages}{033611}, \doiprefix\url{10.1103/PhysRevA.97.033611}
  (\bibinfo{year}{2018}).

\bibitem{Stewart10}
\bibinfo{author}{Stewart, J.~T.}, \bibinfo{author}{Gaebler, J.~P.},
  \bibinfo{author}{Drake, T.~E.} \& \bibinfo{author}{Jin, D.~S.}
\newblock \bibinfo{journal}{\bibinfo{title}{Verification of universal relations
  in a strongly interacting {Fermi} gas}}.
\newblock {\emph{\JournalTitle{Phys. Rev. Lett.}}}
  \textbf{\bibinfo{volume}{104}}, \bibinfo{pages}{235301},
  \doiprefix\url{10.1103/PhysRevLett.104.235301} (\bibinfo{year}{2010}).

\bibitem{Wild12}
\bibinfo{author}{Wild, R.~J.}, \bibinfo{author}{Makotyn, P.},
  \bibinfo{author}{Pino, J.~M.}, \bibinfo{author}{Cornell, E.~A.} \&
  \bibinfo{author}{Jin, D.~S.}
\newblock \bibinfo{journal}{\bibinfo{title}{Measurements of {Tan’s} contact
  in an atomic {Bose-Einstein} condensate}}.
\newblock {\emph{\JournalTitle{Phys. Rev. Lett.}}}
  \textbf{\bibinfo{volume}{108}}, \bibinfo{pages}{145305},
  \doiprefix\url{10.1103/PhysRevLett.108.145305} (\bibinfo{year}{2012}).

\bibitem{Chang16}
\bibinfo{author}{Chang, R.} \emph{et~al.}
\newblock \bibinfo{journal}{\bibinfo{title}{Momentum-resolved observation of
  thermal and quantum depletion in a {Bose} gas}}.
\newblock {\emph{\JournalTitle{Phys. Rev. Lett.}}}
  \textbf{\bibinfo{volume}{117}}, \bibinfo{pages}{235303},
  \doiprefix\url{10.1103/PhysRevLett.117.235303} (\bibinfo{year}{2016}).

\bibitem{Makotyn14}
\bibinfo{author}{Makotyn, P.}, \bibinfo{author}{Klauss, C.~E.},
  \bibinfo{author}{Goldberger, D.~L.}, \bibinfo{author}{Cornell, E.~A.} \&
  \bibinfo{author}{Jin, D.~S.}
\newblock \bibinfo{journal}{\bibinfo{title}{Universal dynamics of a degenerate
  unitary {Bose} gas}}.
\newblock {\emph{\JournalTitle{Nature Physics}}} \textbf{\bibinfo{volume}{10}},
  \bibinfo{pages}{116--119}, \doiprefix\url{10.1038/nphys2850}
  (\bibinfo{year}{2014}).

\bibitem{Eigen18}
\bibinfo{author}{Eigen, C.} \emph{et~al.}
\newblock \bibinfo{journal}{\bibinfo{title}{Universal prethermal dynamics of
  {Bose} gases quenched to unitarity}}.
\newblock {\emph{\JournalTitle{Nature}}} \bibinfo{pages}{221},
  \doiprefix\url{https://doi.org/10.1038/s41586-018-0674-1}
  (\bibinfo{year}{2018}).

\bibitem{Xu06}
\bibinfo{author}{Xu, K.} \emph{et~al.}
\newblock \bibinfo{journal}{\bibinfo{title}{Observation of strong quantum
  depletion in a gaseous {Bose-Einstein} condensate}}.
\newblock {\emph{\JournalTitle{Phys. Rev. Lett.}}}
  \textbf{\bibinfo{volume}{96}}, \bibinfo{pages}{180405},
  \doiprefix\url{10.1103/PhysRevLett.96.180405} (\bibinfo{year}{2006}).

\bibitem{pieczarka20}
\bibinfo{author}{Pieczarka, M.} \emph{et~al.}
\newblock \bibinfo{journal}{\bibinfo{title}{Observation of quantum depletion in
  a non-equilibrium exciton–polariton condensate}}.
\newblock {\emph{\JournalTitle{Nature Communications}}}
  \textbf{\bibinfo{volume}{11}}, \bibinfo{pages}{429},
  \doiprefix\url{10.1038/s41467-019-14243-6} (\bibinfo{year}{2020}).

\bibitem{Lopes17_depletion}
\bibinfo{author}{Lopes, R.} \emph{et~al.}
\newblock \bibinfo{journal}{\bibinfo{title}{Quantum depletion of a homogeneous
  {Bose-Einstein} condensate}}.
\newblock {\emph{\JournalTitle{Phys. Rev. Lett.}}}
  \textbf{\bibinfo{volume}{119}}, \bibinfo{pages}{190404},
  \doiprefix\url{10.1103/PhysRevLett.119.190404} (\bibinfo{year}{2017}).

\bibitem{Cayla20}
\bibinfo{author}{Cayla, H.} \emph{et~al.}
\newblock \bibinfo{journal}{\bibinfo{title}{{Hanbury Brown and Twiss} bunching
  of phonons and of the quantum depletion in an interacting {Bose} gas}}.
\newblock {\emph{\JournalTitle{Phys. Rev. Lett.}}}
  \textbf{\bibinfo{volume}{125}}, \bibinfo{pages}{165301},
  \doiprefix\url{10.1103/PhysRevLett.125.165301} (\bibinfo{year}{2020}).

\bibitem{Kuhnle11}
\bibinfo{author}{Kuhnle, E.~D.} \emph{et~al.}
\newblock \bibinfo{journal}{\bibinfo{title}{Temperature dependence of the
  universal contact parameter in a unitary {Fermi} gas}}.
\newblock {\emph{\JournalTitle{Phys. Rev. Lett.}}}
  \textbf{\bibinfo{volume}{106}}, \bibinfo{pages}{170402},
  \doiprefix\url{10.1103/PhysRevLett.106.170402} (\bibinfo{year}{2011}).

\bibitem{Sagi12}
\bibinfo{author}{Sagi, Y.}, \bibinfo{author}{Drake, T.~E.},
  \bibinfo{author}{Paudel, R.} \& \bibinfo{author}{Jin, D.~S.}
\newblock \bibinfo{journal}{\bibinfo{title}{Measurement of the homogeneous
  contact of a unitary {Fermi} gas}}.
\newblock {\emph{\JournalTitle{Phys. Rev. Lett.}}}
  \textbf{\bibinfo{volume}{109}}, \bibinfo{pages}{220402},
  \doiprefix\url{10.1103/PhysRevLett.109.220402} (\bibinfo{year}{2012}).

\bibitem{Fletcher17}
\bibinfo{author}{Fletcher, R.~J.} \emph{et~al.}
\newblock \bibinfo{journal}{\bibinfo{title}{Two- and three-body contacts in the
  unitary {Bose} gas}}.
\newblock {\emph{\JournalTitle{Science}}} \textbf{\bibinfo{volume}{355}},
  \bibinfo{pages}{377--380}, \doiprefix\url{10.1126/science.aai8195}
  (\bibinfo{year}{2017}).

\bibitem{Lopes17_quasiparticle}
\bibinfo{author}{Lopes, R.} \emph{et~al.}
\newblock \bibinfo{journal}{\bibinfo{title}{Quasiparticle energy in a strongly
  interacting homogeneous {Bose-Einstein} condensate}}.
\newblock {\emph{\JournalTitle{Phys. Rev. Lett.}}}
  \textbf{\bibinfo{volume}{118}}, \bibinfo{pages}{210401},
  \doiprefix\url{10.1103/PhysRevLett.118.210401} (\bibinfo{year}{2017}).

\bibitem{Mukherjee19}
\bibinfo{author}{Mukherjee, B.} \emph{et~al.}
\newblock \bibinfo{journal}{\bibinfo{title}{Spectral response and contact of
  the unitary {Fermi} gas}}.
\newblock {\emph{\JournalTitle{Phys. Rev. Lett.}}}
  \textbf{\bibinfo{volume}{122}}, \bibinfo{pages}{203402},
  \doiprefix\url{10.1103/PhysRevLett.122.203402} (\bibinfo{year}{2019}).

\bibitem{Carcy19}
\bibinfo{author}{Carcy, C.} \emph{et~al.}
\newblock \bibinfo{journal}{\bibinfo{title}{Contact and sum rules in a
  near-uniform {Fermi} gas at unitarity}}.
\newblock {\emph{\JournalTitle{Phys. Rev. Lett.}}}
  \textbf{\bibinfo{volume}{122}}, \bibinfo{pages}{203401},
  \doiprefix\url{10.1103/PhysRevLett.122.203401} (\bibinfo{year}{2019}).

\bibitem{Colussi20}
\bibinfo{author}{Colussi, V.~E.} \emph{et~al.}
\newblock \bibinfo{journal}{\bibinfo{title}{Cumulant theory of the unitary
  {Bose} gas: Prethermal and {Efimovian} dynamics}}.
\newblock {\emph{\JournalTitle{Phys. Rev. A}}} \textbf{\bibinfo{volume}{102}},
  \bibinfo{pages}{063314}, \doiprefix\url{10.1103/PhysRevA.102.063314}
  (\bibinfo{year}{2020}).

\bibitem{Kira15_coherent}
\bibinfo{author}{Kira, M.}
\newblock \bibinfo{journal}{\bibinfo{title}{Coherent quantum depletion of an
  interacting atom condensate}}.
\newblock {\emph{\JournalTitle{Nature Communications}}}
  \textbf{\bibinfo{volume}{6}}, \bibinfo{pages}{6624},
  \doiprefix\url{10.1038/ncomms7624} (\bibinfo{year}{2015}).

\bibitem{Smith14}
\bibinfo{author}{Smith, D.~H.}, \bibinfo{author}{Braaten, E.},
  \bibinfo{author}{Kang, D.} \& \bibinfo{author}{Platter, L.}
\newblock \bibinfo{journal}{\bibinfo{title}{Two-body and three-body contacts
  for identical bosons near unitarity}}.
\newblock {\emph{\JournalTitle{Phys. Rev. Lett.}}}
  \textbf{\bibinfo{volume}{112}}, \bibinfo{pages}{110402},
  \doiprefix\url{10.1103/PhysRevLett.112.110402} (\bibinfo{year}{2014}).

\bibitem{Qu16}
\bibinfo{author}{Qu, C.}, \bibinfo{author}{Pitaevskii, L.~P.} \&
  \bibinfo{author}{Stringari, S.}
\newblock \bibinfo{journal}{\bibinfo{title}{Expansion of harmonically trapped
  interacting particles and time dependence of the contact}}.
\newblock {\emph{\JournalTitle{Phys. Rev. A}}} \textbf{\bibinfo{volume}{94}},
  \bibinfo{pages}{063635}, \doiprefix\url{10.1103/PhysRevA.94.063635}
  (\bibinfo{year}{2016}).

\bibitem{Braaten10}
\bibinfo{author}{Braaten, E.}, \bibinfo{author}{Kang, D.} \&
  \bibinfo{author}{Platter, L.}
\newblock \bibinfo{journal}{\bibinfo{title}{Short-time operator product
  expansion for rf spectroscopy of a strongly interacting {Fermi} gas}}.
\newblock {\emph{\JournalTitle{Phys. Rev. Lett.}}}
  \textbf{\bibinfo{volume}{104}}, \bibinfo{pages}{223004},
  \doiprefix\url{10.1103/PhysRevLett.104.223004} (\bibinfo{year}{2010}).

\bibitem{Braaten11}
\bibinfo{author}{Braaten, E.}, \bibinfo{author}{Kang, D.} \&
  \bibinfo{author}{Platter, L.}
\newblock \bibinfo{journal}{\bibinfo{title}{Universal relations for identical
  bosons from three-body physics}}.
\newblock {\emph{\JournalTitle{Phys. Rev. Lett.}}}
  \textbf{\bibinfo{volume}{106}}, \bibinfo{pages}{153005},
  \doiprefix\url{10.1103/PhysRevLett.106.153005} (\bibinfo{year}{2011}).

\bibitem{Rakhimov20}
\bibinfo{author}{Rakhimov, A.}
\newblock \bibinfo{journal}{\bibinfo{title}{{Tan's} contact as an indicator of
  completeness and self-consistency of a theory}}.
\newblock {\emph{\JournalTitle{Phys. Rev. A}}} \textbf{\bibinfo{volume}{102}},
  \bibinfo{pages}{063306}, \doiprefix\url{10.1103/PhysRevA.102.063306}
  (\bibinfo{year}{2020}).

\bibitem{Braaten08}
\bibinfo{author}{Braaten, E.} \& \bibinfo{author}{Platter, L.}
\newblock \bibinfo{journal}{\bibinfo{title}{Exact relations for a strongly
  interacting {Fermi} gas from the operator product expansion}}.
\newblock {\emph{\JournalTitle{Phys. Rev. Lett.}}}
  \textbf{\bibinfo{volume}{100}}, \bibinfo{pages}{205301},
  \doiprefix\url{10.1103/PhysRevLett.100.205301} (\bibinfo{year}{2008}).

\bibitem{Zhang09}
\bibinfo{author}{Zhang, S.} \& \bibinfo{author}{Leggett, A.~J.}
\newblock \bibinfo{journal}{\bibinfo{title}{Universal properties of the
  ultracold {Fermi} gas}}.
\newblock {\emph{\JournalTitle{Phys. Rev. A}}} \textbf{\bibinfo{volume}{79}},
  \bibinfo{pages}{023601}, \doiprefix\url{10.1103/physreva.79.023601}
  (\bibinfo{year}{2009}).

\bibitem{Combescot09}
\bibinfo{author}{Combescot, R.}, \bibinfo{author}{Alzetto, F.} \&
  \bibinfo{author}{Leyronas, X.}
\newblock \bibinfo{journal}{\bibinfo{title}{Particle distribution tail and
  related energy formula}}.
\newblock {\emph{\JournalTitle{Phys. Rev. A}}} \textbf{\bibinfo{volume}{79}},
  \bibinfo{pages}{053640}, \doiprefix\url{10.1103/PhysRevA.79.053640}
  (\bibinfo{year}{2009}).

\bibitem{Werner12_boson}
\bibinfo{author}{Werner, F.} \& \bibinfo{author}{Castin, Y.}
\newblock \bibinfo{journal}{\bibinfo{title}{General relations for quantum gases
  in two and three dimensions. {II}. {Bosons} and mixtures}}.
\newblock {\emph{\JournalTitle{Phys. Rev. A}}} \textbf{\bibinfo{volume}{86}},
  \bibinfo{pages}{053633}, \doiprefix\url{10.1103/PhysRevA.86.053633}
  (\bibinfo{year}{2012}).

\bibitem{Werner12_fermion}
\bibinfo{author}{Werner, F.} \& \bibinfo{author}{Castin, Y.}
\newblock \bibinfo{journal}{\bibinfo{title}{General relations for quantum gases
  in two and three dimensions: {Two}-component fermions}}.
\newblock {\emph{\JournalTitle{Phys. Rev. A}}} \textbf{\bibinfo{volume}{86}},
  \bibinfo{pages}{013626}, \doiprefix\url{10.1103/PhysRevA.86.013626}
  (\bibinfo{year}{2012}).

\bibitem{Sinatra00}
\bibinfo{author}{Sinatra, A.}, \bibinfo{author}{Castin, Y.} \&
  \bibinfo{author}{Lobo, C.}
\newblock \bibinfo{journal}{\bibinfo{title}{A {Monte Carlo} formulation of the
  {Bogolubov} theory}}.
\newblock {\emph{\JournalTitle{Journal of Modern Optics}}}
  \textbf{\bibinfo{volume}{47}}, \bibinfo{pages}{2629--2644},
  \doiprefix\url{10.1080/09500340008232186} (\bibinfo{year}{2000}).

\bibitem{Deuar11}
\bibinfo{author}{Deuar, P.}, \bibinfo{author}{Chwede\'{n}czuk, J.},
  \bibinfo{author}{Trippenbach, M.} \& \bibinfo{author}{Zi\'{n}, P.}
\newblock \bibinfo{journal}{\bibinfo{title}{{Bogoliubov} dynamics of condensate
  collisions using the positive-{P} representation}}.
\newblock {\emph{\JournalTitle{Phys. Rev. A}}} \textbf{\bibinfo{volume}{83}},
  \bibinfo{pages}{063625}, \doiprefix\url{10.1103/PhysRevA.83.063625}
  (\bibinfo{year}{2011}).

\bibitem{Steinhauer03}
\bibinfo{author}{Steinhauer, J.} \emph{et~al.}
\newblock \bibinfo{journal}{\bibinfo{title}{{Bragg} spectroscopy of the
  multibranch {Bogoliubov} spectrum of elongated {Bose-Einstein} condensates}}.
\newblock {\emph{\JournalTitle{Phys. Rev. Lett.}}}
  \textbf{\bibinfo{volume}{90}}, \bibinfo{pages}{060404},
  \doiprefix\url{10.1103/PhysRevLett.90.060404} (\bibinfo{year}{2003}).

\bibitem{Cayla22}
\bibinfo{author}{Cayla, H.} \emph{et~al.}
\newblock \bibinfo{journal}{\bibinfo{title}{Observation of $1/k^4$-tails in the
  asymptotic momentum distribution of bose polarons}}.
\newblock {\emph{\JournalTitle{arXiv: 2204.10697}}}
  \doiprefix\url{10.48550/arXiv.2204.10697} (\bibinfo{year}{2022}).

\bibitem{Dmowski17}
\bibinfo{author}{Dmowski, W.} \emph{et~al.}
\newblock \bibinfo{journal}{\bibinfo{title}{Observation of dynamic atom-atom
  correlation in liquid helium in real space}}.
\newblock {\emph{\JournalTitle{Nature Communications}}}
  \textbf{\bibinfo{volume}{8}}, \bibinfo{pages}{15294}, \doiprefix\url{doi:
  10.1038/ncomms15294} (\bibinfo{year}{2017}).

\bibitem{Glyde00}
\bibinfo{author}{Glyde, H.~R.}, \bibinfo{author}{Azuah, R.~T.} \&
  \bibinfo{author}{Stirling, W.~G.}
\newblock \bibinfo{journal}{\bibinfo{title}{Condensate, momentum distribution,
  and final-state effects in liquid ${}^{4}\mathrm{He}$}}.
\newblock {\emph{\JournalTitle{Phys. Rev. B}}} \textbf{\bibinfo{volume}{62}},
  \bibinfo{pages}{14337--14349}, \doiprefix\url{10.1103/PhysRevB.62.14337}
  (\bibinfo{year}{2000}).

\bibitem{Moroni04}
\bibinfo{author}{Moroni, S.} \& \bibinfo{author}{Boninsegni, M.}
\newblock \bibinfo{journal}{\bibinfo{title}{Condensate fraction in liquid
  $^4${He}}}.
\newblock {\emph{\JournalTitle{Journal of Low Temperature Physics}}}
  \textbf{\bibinfo{volume}{136}}, \bibinfo{pages}{129},
  \doiprefix\url{https://doi.org/10.1023/B:JOLT.0000038518.10132.30}
  (\bibinfo{year}{2004}).

\bibitem{Tan08_momentum}
\bibinfo{author}{Tan, S.}
\newblock \bibinfo{journal}{\bibinfo{title}{Large momentum part of a strongly
  correlated {Fermi} gas}}.
\newblock {\emph{\JournalTitle{Annals of Physics}}}
  \textbf{\bibinfo{volume}{323}}, \bibinfo{pages}{2971--2986},
  \doiprefix\url{10.1016/j.aop.2008.03.005} (\bibinfo{year}{2008}).

\bibitem{Tan08_energetics}
\bibinfo{author}{Tan, S.}
\newblock \bibinfo{journal}{\bibinfo{title}{Energetics of a strongly correlated
  {Fermi} gas}}.
\newblock {\emph{\JournalTitle{Annals of Physics}}}
  \textbf{\bibinfo{volume}{323}}, \bibinfo{pages}{2952--2970},
  \doiprefix\url{10.1016/j.aop.2008.03.004} (\bibinfo{year}{2008}).

\bibitem{Tan08_virial}
\bibinfo{author}{Tan, S.}
\newblock \bibinfo{journal}{\bibinfo{title}{Generalized virial theorem and
  pressure relation for a strongly correlated {Fermi} gas}}.
\newblock {\emph{\JournalTitle{Annals of Physics}}}
  \textbf{\bibinfo{volume}{323}}, \bibinfo{pages}{2987--2990},
  \doiprefix\url{10.1016/j.aop.2008.03.003} (\bibinfo{year}{2008}).

\bibitem{Hoinka15}
\bibinfo{author}{Hoinka, S.} \emph{et~al.}
\newblock \bibinfo{journal}{\bibinfo{title}{Precise determination of the
  structure factor and contact in a unitary {Fermi} gas}}.
\newblock {\emph{\JournalTitle{Phys. Rev. Lett.}}}
  \textbf{\bibinfo{volume}{110}}, \bibinfo{pages}{055305},
  \doiprefix\url{10.1103/PhysRevLett.110.055305} (\bibinfo{year}{2013}).

\bibitem{Tenart21}
\bibinfo{author}{Tenart, A.}, \bibinfo{author}{Herc\'{e}, G.},
  \bibinfo{author}{Bureik, J.}, \bibinfo{author}{Dareau, A.} \&
  \bibinfo{author}{Cl\'{e}ment, D.}
\newblock \bibinfo{journal}{\bibinfo{title}{Observation of pairs of atoms at
  opposite momenta in an equilibrium interacting bose gas}}.
\newblock {\emph{\JournalTitle{Nature Physics}}} \textbf{\bibinfo{volume}{17}},
  \doiprefix\url{https://doi.org/10.1038/s41567-021-01381-2}
  (\bibinfo{year}{2021}).

\bibitem{Kheruntsyan12}
\bibinfo{author}{Kheruntsyan, K.~V.} \emph{et~al.}
\newblock \bibinfo{journal}{\bibinfo{title}{Violation of the {Cauchy-Schwarz}
  inequality with matter waves}}.
\newblock {\emph{\JournalTitle{Phys. Rev. Lett.}}}
  \textbf{\bibinfo{volume}{108}}, \bibinfo{pages}{260401},
  \doiprefix\url{10.1103/PhysRevLett.108.260401} (\bibinfo{year}{2012}).

\bibitem{PethickSmith}
\bibinfo{author}{Pethick, C.} \& \bibinfo{author}{Smith, H.}
\newblock \emph{\bibinfo{title}{{Bose-Einstein} condensation in dilute gases}}
  (\bibinfo{publisher}{Cambridge University Press},
  \bibinfo{address}{Cambridge, New York}, \bibinfo{year}{2008}),
  \bibinfo{edition}{2nd ed} edn.

\bibitem{Olshanii03}
\bibinfo{author}{Olshanii, M.} \& \bibinfo{author}{Dunjko, V.}
\newblock \bibinfo{journal}{\bibinfo{title}{Short-distance correlation
  properties of the {Lieb-Liniger} system and momentum distributions of trapped
  one-dimensional atomic gases}}.
\newblock {\emph{\JournalTitle{Phys. Rev. Lett.}}}
  \textbf{\bibinfo{volume}{91}}, \bibinfo{pages}{090401},
  \doiprefix\url{10.1103/PhysRevLett.91.090401} (\bibinfo{year}{2003}).

\bibitem{Dall07}
\bibinfo{author}{Dall, R.} \& \bibinfo{author}{Truscott, A.}
\newblock \bibinfo{journal}{\bibinfo{title}{{Bose–Einstein} condensation of
  metastable helium in a bi-planar quadrupole {Ioffe} configuration trap}}.
\newblock {\emph{\JournalTitle{Optics Communications}}}
  \textbf{\bibinfo{volume}{270}}, \bibinfo{pages}{255–261},
  \doiprefix\url{10.1016/j.optcom.2006.09.031} (\bibinfo{year}{2007}).

\bibitem{Moal06}
\bibinfo{author}{Moal, S.} \emph{et~al.}
\newblock \bibinfo{journal}{\bibinfo{title}{Accurate determination of the
  scattering length of metastable helium atoms using dark resonances between
  atoms and exotic molecules}}.
\newblock {\emph{\JournalTitle{Phys. Rev. Lett.}}}
  \textbf{\bibinfo{volume}{96}}, \doiprefix\url{10.1103/physrevlett.96.023203}
  (\bibinfo{year}{2006}).

\bibitem{Manning10}
\bibinfo{author}{Manning, A.~G.}, \bibinfo{author}{Hodgman, S.~S.},
  \bibinfo{author}{Dall, R.~G.}, \bibinfo{author}{Johnsson, M.~T.} \&
  \bibinfo{author}{Truscott, A.~G.}
\newblock \bibinfo{journal}{\bibinfo{title}{The {Hanbury Brown-Twiss} effect in
  a pulsed atom laser}}.
\newblock {\emph{\JournalTitle{Optics Express}}} \textbf{\bibinfo{volume}{18}},
  \bibinfo{pages}{18712}, \doiprefix\url{10.1364/oe.18.018712}
  (\bibinfo{year}{2010}).

\bibitem{Hodgman09}
\bibinfo{author}{Hodgman, S.~S.} \emph{et~al.}
\newblock \bibinfo{journal}{\bibinfo{title}{Metastable helium: A new
  determination of the longest atomic excited-state lifetime}}.
\newblock {\emph{\JournalTitle{Phys. Rev. Lett.}}}
  \textbf{\bibinfo{volume}{103}}, \bibinfo{pages}{053002},
  \doiprefix\url{10.1103/physrevlett.103.053002} (\bibinfo{year}{2009}).

\bibitem{Hodgman11}
\bibinfo{author}{Hodgman, S.~S.}, \bibinfo{author}{Dall, R.~G.},
  \bibinfo{author}{Manning, A.~G.}, \bibinfo{author}{Baldwin, K. G.~H.} \&
  \bibinfo{author}{Truscott, A.~G.}
\newblock \bibinfo{journal}{\bibinfo{title}{Direct measurement of long-range
  third-order coherence in {Bose-Einstein} condensates}}.
\newblock {\emph{\JournalTitle{Science}}} \bibinfo{pages}{1046--1049}
  (\bibinfo{year}{2011}).

\bibitem{Dall13}
\bibinfo{author}{Dall, R.~G.} \emph{et~al.}
\newblock \bibinfo{journal}{\bibinfo{title}{Ideal \emph{n}-body correlations
  with massive particles}}.
\newblock {\emph{\JournalTitle{Nature physics}}} \bibinfo{pages}{341--344}
  (\bibinfo{year}{2013}).

\bibitem{Schellekens05}
\bibinfo{author}{Schellekens, M.} \emph{et~al.}
\newblock \bibinfo{journal}{\bibinfo{title}{{Hanbury Brown Twiss} effect for
  ultracold quantum gases}}.
\newblock {\emph{\JournalTitle{Science}}} \bibinfo{pages}{648--651}
  (\bibinfo{year}{2005}).

\bibitem{Jeltes07}
\bibinfo{author}{Jeltes, T.} \emph{et~al.}
\newblock \bibinfo{journal}{\bibinfo{title}{Comparison of the {Hanbury
  Brown–Twiss} effect for bosons and fermions}}.
\newblock {\emph{\JournalTitle{Nature}}} \bibinfo{pages}{402--405}
  (\bibinfo{year}{2007}).

\bibitem{Dall11}
\bibinfo{author}{Dall, R.~G.} \emph{et~al.}
\newblock \bibinfo{journal}{\bibinfo{title}{Observation of atomic speckle and
  {Hanbury Brown–Twiss} correlations in guided matter waves}}.
\newblock {\emph{\JournalTitle{Nature communications}}}
  (\bibinfo{year}{2011}).

\bibitem{Perrin07}
\bibinfo{author}{Perrin, A.} \emph{et~al.}
\newblock \bibinfo{journal}{\bibinfo{title}{Observation of atom pairs in
  spontaneous four-wave mixing of two colliding {Bose-Einstein} condensates}}.
\newblock {\emph{\JournalTitle{Phys. Rev. Lett.}}}
  \textbf{\bibinfo{volume}{99}}, \bibinfo{pages}{150405},
  \doiprefix\url{10.1103/PhysRevLett.99.150405} (\bibinfo{year}{2007}).

\bibitem{Perrin12}
\bibinfo{author}{A, P.} \emph{et~al.}
\newblock \bibinfo{journal}{\bibinfo{title}{{Hanbury Brown and Twiss}
  correlations across the {Bose-Einstein} condensation threshold}}.
\newblock {\emph{\JournalTitle{Nature physics}}} \bibinfo{pages}{195--198}
  (\bibinfo{year}{2012}).

\bibitem{Dalfovo99}
\bibinfo{author}{Dalfovo, F.}, \bibinfo{author}{Giorgini, S.},
  \bibinfo{author}{Pitaevskii, L.~P.} \& \bibinfo{author}{Stringari, S.}
\newblock \bibinfo{journal}{\bibinfo{title}{Theory of bose-einstein
  condensation in trapped gases}}.
\newblock {\emph{\JournalTitle{Rev. Mod. Phys.}}}
  \textbf{\bibinfo{volume}{71}}, \bibinfo{pages}{463--512},
  \doiprefix\url{10.1103/RevModPhys.71.463} (\bibinfo{year}{1999}).

\bibitem{Clauset09}
\bibinfo{author}{Clauset, A.}, \bibinfo{author}{Shalizi, C.~R.} \&
  \bibinfo{author}{Newman, M. E.~J.}
\newblock \bibinfo{journal}{\bibinfo{title}{Power-law distributions in
  empirical data}}.
\newblock {\emph{\JournalTitle{SIAM Review}}} \textbf{\bibinfo{volume}{51}},
  \bibinfo{pages}{661--703}, \doiprefix\url{10.1137/070710111}
  (\bibinfo{year}{2009}).
\newblock \bibinfo{note}{ArXiv: 0706.1062}.

\bibitem{Virkar14}
\bibinfo{author}{Virkar, Y.} \& \bibinfo{author}{Clauset, A.}
\newblock \bibinfo{journal}{\bibinfo{title}{Power-law distributions in binned
  empirical data}}.
\newblock {\emph{\JournalTitle{The Annals of Applied Statistics}}}
  \textbf{\bibinfo{volume}{8}}, \bibinfo{pages}{89--119},
  \doiprefix\url{10.1214/13-AOAS710} (\bibinfo{year}{2014}).
\newblock \bibinfo{note}{ArXiv: 1208.3524}.

\bibitem{stob}
\bibinfo{author}{Pietraszewicz, J.},
  \bibinfo{author}{Stobi\ifmmode~\acute{n}\else \'{n}\fi{}ska, M.} \&
  \bibinfo{author}{Deuar, P.}
\newblock \bibinfo{journal}{\bibinfo{title}{Correlation evolution in dilute
  {Bose-Einstein} condensates after quantum quenches}}.
\newblock {\emph{\JournalTitle{Phys. Rev. A}}} \textbf{\bibinfo{volume}{99}},
  \bibinfo{pages}{023620}, \doiprefix\url{10.1103/PhysRevA.99.023620}
  (\bibinfo{year}{2019}).

\bibitem{Vogels02b}
\bibinfo{author}{Vogels, J.~M.}, \bibinfo{author}{Xu, K.} \&
  \bibinfo{author}{Ketterle, W.}
\newblock \bibinfo{journal}{\bibinfo{title}{Generation of macroscopic
  pair-correlated atomic beams by four-wave mixing in bose-einstein
  condensates}}.
\newblock {\emph{\JournalTitle{Phys. Rev. Lett.}}}
  \textbf{\bibinfo{volume}{89}}, \bibinfo{pages}{020401},
  \doiprefix\url{10.1103/PhysRevLett.89.020401} (\bibinfo{year}{2002}).

\bibitem{Krachmalnicoff10}
\bibinfo{author}{Krachmalnicoff, V.} \emph{et~al.}
\newblock \bibinfo{journal}{\bibinfo{title}{Spontaneous four-wave mixing of de
  {Broglie} waves: Beyond optics}}.
\newblock {\emph{\JournalTitle{Phys. Rev. Lett.}}}
  \textbf{\bibinfo{volume}{104}}, \bibinfo{pages}{150402},
  \doiprefix\url{10.1103/PhysRevLett.104.150402} (\bibinfo{year}{2010}).

\bibitem{Deuar11b}
\bibinfo{author}{Deuar, P.}, \bibinfo{author}{Zi\'n, P.},
  \bibinfo{author}{Chwede\'nczuk, J.} \& \bibinfo{author}{Trippenbach, M.}
\newblock \bibinfo{journal}{\bibinfo{title}{Mean field effects on the scattered
  atoms in condensate collisions}}.
\newblock {\emph{\JournalTitle{The European Physical Journal D}}}
  \textbf{\bibinfo{volume}{65}}, \bibinfo{pages}{19--24},
  \doiprefix\url{10.1140/epjd/e2011-20066-7} (\bibinfo{year}{2011}).

\bibitem{Ozeri02}
\bibinfo{author}{Ozeri, R.}, \bibinfo{author}{Steinhauer, J.},
  \bibinfo{author}{Katz, N.} \& \bibinfo{author}{Davidson, N.}
\newblock \bibinfo{journal}{\bibinfo{title}{Direct observation of the phonon
  energy in a bose-einstein condensate by tomographic imaging}}.
\newblock {\emph{\JournalTitle{Phys. Rev. Lett.}}}
  \textbf{\bibinfo{volume}{88}}, \bibinfo{pages}{220401},
  \doiprefix\url{10.1103/PhysRevLett.88.220401} (\bibinfo{year}{2002}).

\bibitem{Simsarian00}
\bibinfo{author}{Simsarian, J.~E.} \emph{et~al.}
\newblock \bibinfo{journal}{\bibinfo{title}{Imaging the phase of an evolving
  bose-einstein condensate wave function}}.
\newblock {\emph{\JournalTitle{Phys. Rev. Lett.}}}
  \textbf{\bibinfo{volume}{85}}, \bibinfo{pages}{2040--2043},
  \doiprefix\url{10.1103/PhysRevLett.85.2040} (\bibinfo{year}{2000}).

\bibitem{Deuar14}
\bibinfo{author}{Deuar, P.} \emph{et~al.}
\newblock \bibinfo{journal}{\bibinfo{title}{Anisotropy in s -wave
  {Bose}-{Einstein} condensate collisions and its relationship to
  superradiance}}.
\newblock {\emph{\JournalTitle{Phys. Rev. A}}} \textbf{\bibinfo{volume}{90}},
  \bibinfo{pages}{033613} (\bibinfo{year}{2014}).

\bibitem{Hodgman17}
\bibinfo{author}{Hodgman, S.~S.}, \bibinfo{author}{Khakimov, R.~I.},
  \bibinfo{author}{Lewis-Swan, R.~J.}, \bibinfo{author}{Truscott, A.~G.} \&
  \bibinfo{author}{Kheruntsyan, K.~V.}
\newblock \bibinfo{journal}{\bibinfo{title}{Solving the quantum many-body
  problem via correlations measured with a momentum microscope}}.
\newblock {\emph{\JournalTitle{Phys. Rev. Lett.}}}
  \textbf{\bibinfo{volume}{118}}, \bibinfo{pages}{240402},
  \doiprefix\url{10.1103/PhysRevLett.118.240402} (\bibinfo{year}{2017}).

\bibitem{Vassen16}
\bibinfo{author}{Vassen, W.}, \bibinfo{author}{Notermans, R. P. M. J.~W.},
  \bibinfo{author}{Rengelink, R.~J.} \& \bibinfo{author}{van~der Beek, R. F.
  H.~J.}
\newblock \bibinfo{journal}{\bibinfo{title}{Ultracold metastable helium:
  {Ramsey} fringes and atom interferometry}}.
\newblock {\emph{\JournalTitle{Applied Physics B}}}
  \textbf{\bibinfo{volume}{122}}, \bibinfo{pages}{289},
  \doiprefix\url{10.1007/s00340-016-6563-0} (\bibinfo{year}{2016}).

\bibitem{Dedman07}
\bibinfo{author}{Dedman, C.~J.}, \bibinfo{author}{Dall, R.~G.},
  \bibinfo{author}{Byron, L.~J.} \& \bibinfo{author}{Truscott, A.~G.}
\newblock \bibinfo{journal}{\bibinfo{title}{Active cancellation of stray
  magnetic fields in a {Bose-Einstein} condensation experiment}}.
\newblock {\emph{\JournalTitle{Review of Scientific Instruments}}}
  \textbf{\bibinfo{volume}{78}}, \bibinfo{pages}{024703},
  \doiprefix\url{10.1063/1.2472600} (\bibinfo{year}{2007}).

\bibitem{Henson18}
\bibinfo{author}{Henson, B.~M.} \emph{et~al.}
\newblock \bibinfo{journal}{\bibinfo{title}{{Bogoliubov-Cherenkov} radiation in
  an atom laser}}.
\newblock {\emph{\JournalTitle{Phys. Rev. A}}} \textbf{\bibinfo{volume}{97}},
  \bibinfo{pages}{063601}, \doiprefix\url{10.1103/physreva.97.063601}
  (\bibinfo{year}{2018}).

\bibitem{shin19}
\bibinfo{author}{Shin, D.~K.} \emph{et~al.}
\newblock \bibinfo{journal}{\bibinfo{title}{{Bell} correlations between
  spatially separated pairs of atoms}}.
\newblock {\emph{\JournalTitle{Nature communications}}} \bibinfo{pages}{4447},
  \doiprefix\url{https://doi.org/10.1038/s41467-019-12192-8}
  (\bibinfo{year}{2019}).

\bibitem{shin20}
\bibinfo{author}{Shin, D.~K.}, \bibinfo{author}{Ross, J.~A.},
  \bibinfo{author}{Henson, B.~M.}, \bibinfo{author}{Hodgman, S.~S.} \&
  \bibinfo{author}{Truscott, A.~G.}
\newblock \bibinfo{journal}{\bibinfo{title}{Entanglement-base {3D} magnetic
  gradiometry with an ultracold atomic scattering halo}}.
\newblock {\emph{\JournalTitle{New Journal of Physics}}}
  \bibinfo{pages}{013002}, \doiprefix\url{10.1088/1367-2630/ab66de}
  (\bibinfo{year}{2020}).

\bibitem{Jaskula10}
\bibinfo{author}{Jaskula, J.-C.} \emph{et~al.}
\newblock \bibinfo{journal}{\bibinfo{title}{Sub-{Poissonian} number differences
  in four-wave mixing of matter waves}}.
\newblock {\emph{\JournalTitle{Phys. Rev. Lett.}}}
  \textbf{\bibinfo{volume}{105}}, \bibinfo{pages}{190402},
  \doiprefix\url{10.1103/PhysRevLett.105.190402} (\bibinfo{year}{2010}).

\bibitem{Drummond80}
\bibinfo{author}{Drummond, P.~D.} \& \bibinfo{author}{Gardiner, C.~W.}
\newblock \bibinfo{journal}{\bibinfo{title}{Generalised {P}-representations in
  quantum optics}}.
\newblock {\emph{\JournalTitle{Journal of Physics A: Mathematical and
  General}}} \textbf{\bibinfo{volume}{13}}, \bibinfo{pages}{2353}
  (\bibinfo{year}{1980}).

\bibitem{Deuar07}
\bibinfo{author}{Deuar, P.} \& \bibinfo{author}{Drummond, P.~D.}
\newblock \bibinfo{journal}{\bibinfo{title}{Correlations in a {BEC} collision:
  First-principles quantum dynamics with 150\,000 atoms}}.
\newblock {\emph{\JournalTitle{Phys. Rev. Lett.}}}
  \textbf{\bibinfo{volume}{98}}, \bibinfo{pages}{120402},
  \doiprefix\url{10.1103/PhysRevLett.98.120402} (\bibinfo{year}{2007}).

\bibitem{Lewis-Swan14}
\bibinfo{author}{Lewis-Swan, R.~J.} \& \bibinfo{author}{Kheruntsyan, K.~V.}
\newblock \bibinfo{journal}{\bibinfo{title}{Proposal for demonstrating the
  {Hong–Ou–Mandel} effect with matter waves}}.
\newblock {\emph{\JournalTitle{Nature Commun.}}} \textbf{\bibinfo{volume}{5}},
  \bibinfo{pages}{3752}, \doiprefix\url{10.1038/ncomms4752}
  (\bibinfo{year}{2014}).

\bibitem{Lewis-Swan15}
\bibinfo{author}{Lewis-Swan, R.~J.} \& \bibinfo{author}{Kheruntsyan, K.~V.}
\newblock \bibinfo{journal}{\bibinfo{title}{Proposal for a motional-state
  {Bell} inequality test with ultracold atoms}}.
\newblock {\emph{\JournalTitle{Phys. Rev. A}}} \textbf{\bibinfo{volume}{91}},
  \bibinfo{pages}{052114}, \doiprefix\url{10.1103/PhysRevA.91.052114}
  (\bibinfo{year}{2015}).

\bibitem{Deuar13}
\bibinfo{author}{Deuar, P.}, \bibinfo{author}{Wasak, T.},
  \bibinfo{author}{Zi\'{n}, P.}, \bibinfo{author}{Chwede\'{n}czuk, J.} \&
  \bibinfo{author}{Trippenbach, M.}
\newblock \bibinfo{journal}{\bibinfo{title}{Tradeoffs for number squeezing in
  collisions of {Bose-Einstein} condensates}}.
\newblock {\emph{\JournalTitle{Phys. Rev. A}}} \textbf{\bibinfo{volume}{88}},
  \bibinfo{pages}{013617}, \doiprefix\url{10.1103/PhysRevA.88.013617}
  (\bibinfo{year}{2013}).

\bibitem{nstab-longpaper}
\bibinfo{author}{Deuar, P.}, \bibinfo{author}{Ross, J.~D.} \&
  \bibinfo{author}{Truscott, A.~G.} (\bibinfo{year}{2021}).
\newblock \bibinfo{note}{In preparation}.

\bibitem{Castin98}
\bibinfo{author}{Castin, Y.} \& \bibinfo{author}{Dum, R.}
\newblock \bibinfo{journal}{\bibinfo{title}{Low-temperature {Bose-Einstein}
  condensates in time-dependent traps: Beyond the u(1) symmetry-breaking
  approach}}.
\newblock {\emph{\JournalTitle{Phys. Rev. A}}} \textbf{\bibinfo{volume}{57}},
  \bibinfo{pages}{3008--3021},
  \doiprefix\url{https://doi.org/10.1103/PhysRevA.57.3008}
  (\bibinfo{year}{1998}).

\bibitem{tcorr}
\bibinfo{author}{Deuar, P.}
\newblock \bibinfo{journal}{\bibinfo{title}{Multi-time correlations in the
  positive-{P}, {Q}, and doubled phase-space representations}}.
\newblock {\emph{\JournalTitle{{Quantum}}}} \textbf{\bibinfo{volume}{5}},
  \bibinfo{pages}{455},
  \doiprefix\url{https://doi.org/10.22331/q-2021-05-10-455}
  (\bibinfo{year}{2021}).

\bibitem{henson18ML}
\bibinfo{author}{Henson, B.~M.} \emph{et~al.}
\newblock \bibinfo{journal}{\bibinfo{title}{Approaching the adiabatic timescale
  with machine learning}}.
\newblock {\emph{\JournalTitle{Proceedings of the National Academy of
  Sciences}}} \textbf{\bibinfo{volume}{115}}, \bibinfo{pages}{13216},
  \doiprefix\url{https://doi.org/10.1073/pnas.1811501115}
  (\bibinfo{year}{2018}).

\bibitem{Goldstein04}
\bibinfo{author}{Goldstein, M.~L.}, \bibinfo{author}{Morris, S.~A.} \&
  \bibinfo{author}{Yen, G.~G.}
\newblock \bibinfo{journal}{\bibinfo{title}{Problems with fitting to the
  power-law distribution}}.
\newblock {\emph{\JournalTitle{The European Physical Journal B}}}
  \textbf{\bibinfo{volume}{41}}, \bibinfo{pages}{255--258},
  \doiprefix\url{10.1140/epjb/e2004-00316-5} (\bibinfo{year}{2004}).

\bibitem{Hanel17}
\bibinfo{author}{Hanel, R.}, \bibinfo{author}{Corominas-Murtra, B.},
  \bibinfo{author}{Liu, B.} \& \bibinfo{author}{Thurner, S.}
\newblock \bibinfo{journal}{\bibinfo{title}{Fitting power-laws in empirical
  data with estimators that work for all exponents}}.
\newblock {\emph{\JournalTitle{PLOS ONE}}} \textbf{\bibinfo{volume}{12}},
  \bibinfo{pages}{e0170920}, \doiprefix\url{10.1371/journal.pone.0170920}
  (\bibinfo{year}{2017}).

\bibitem{FINESS-Book-Ruostekoski}
\bibinfo{author}{Ruostekoski, J.} \& \bibinfo{author}{Martin, A.~D.}
\newblock \emph{\bibinfo{title}{The Truncated {Wigner} Method for {Bose}
  Gases}}, chap.~\bibinfo{chapter}{13}, \bibinfo{pages}{203--214}
  (\bibinfo{publisher}{Imperial College Press}, \bibinfo{year}{2013}).

\bibitem{FINESS-Book-Sinatra}
\bibinfo{author}{Sinatra, A.}, \bibinfo{author}{Castin, Y.},
  \bibinfo{author}{Carusotto, I.}, \bibinfo{author}{Lobo, C.} \&
  \bibinfo{author}{Witkowska, E.}
\newblock \emph{\bibinfo{title}{Number-Conserving Stochastic Approaches for
  Equilibrium and Time-Dependent {Bose} Gases}}, chap.~\bibinfo{chapter}{14},
  \bibinfo{pages}{215--228} (\bibinfo{publisher}{Imperial College Press},
  \bibinfo{year}{2013}).

\bibitem{Martin10a}
\bibinfo{author}{Martin, A.~D.} \& \bibinfo{author}{Ruostekoski, J.}
\newblock \bibinfo{journal}{\bibinfo{title}{Nonequilibrium quantum dynamics of
  atomic dark solitons}}.
\newblock {\emph{\JournalTitle{New Journal of Physics}}}
  \textbf{\bibinfo{volume}{12}}, \bibinfo{pages}{055018}
  (\bibinfo{year}{2010}).

\bibitem{Martin10b}
\bibinfo{author}{Martin, A.~D.} \& \bibinfo{author}{Ruostekoski, J.}
\newblock \bibinfo{journal}{\bibinfo{title}{Quantum and thermal effects of dark
  solitons in a one-dimensional {Bose} gas}}.
\newblock {\emph{\JournalTitle{Phys. Rev. Lett.}}}
  \textbf{\bibinfo{volume}{104}}, \bibinfo{pages}{194102},
  \doiprefix\url{10.1103/PhysRevLett.104.194102} (\bibinfo{year}{2010}).

\bibitem{Sinatra02}
\bibinfo{author}{Sinatra, A.}, \bibinfo{author}{Lobo, C.} \&
  \bibinfo{author}{Castin, Y.}
\newblock \bibinfo{journal}{\bibinfo{title}{The truncated {Wigner} method for
  {Bose}-condensed gases: limits of validity and applications}}.
\newblock {\emph{\JournalTitle{Journal of Physics B: Atomic, Molecular and
  Optical Physics}}} \textbf{\bibinfo{volume}{35}}, \bibinfo{pages}{3599}
  (\bibinfo{year}{2002}).

\bibitem{Norrie06}
\bibinfo{author}{Norrie, A.~A.}, \bibinfo{author}{Ballagh, R.~J.} \&
  \bibinfo{author}{Gardiner, C.~W.}
\newblock \bibinfo{journal}{\bibinfo{title}{Quantum turbulence and correlations
  in {Bose-Einstein} condensate collisions}}.
\newblock {\emph{\JournalTitle{Phys. Rev. A}}} \textbf{\bibinfo{volume}{73}},
  \bibinfo{pages}{043617}, \doiprefix\url{10.1103/PhysRevA.73.043617}
  (\bibinfo{year}{2006}).

\bibitem{DeuarPhD}
\bibinfo{author}{Deuar, P.}
\newblock \emph{\bibinfo{title}{First-principles quantum simulations of
  many-mode open interacting {Bose} gases using stochastic gauge methods}}.
\newblock Ph.D. thesis, \bibinfo{school}{University of Queensland,
  arXiv:cond-mat/0507023} (\bibinfo{year}{2005}).

\bibitem{Drummond20}
\bibinfo{author}{Drummond, P.~D.} \& \bibinfo{author}{Opanchuk, B.}
\newblock \bibinfo{journal}{\bibinfo{title}{Initial states for quantum field
  simulations in phase space}}.
\newblock {\emph{\JournalTitle{Phys. Rev. Research}}}
  \textbf{\bibinfo{volume}{2}}, \bibinfo{pages}{033304},
  \doiprefix\url{10.1103/PhysRevResearch.2.033304} (\bibinfo{year}{2020}).

\bibitem{Castin96}
\bibinfo{author}{Castin, Y.} \& \bibinfo{author}{Dum, R.}
\newblock \bibinfo{journal}{\bibinfo{title}{{Bose-Einstein} condensates in time
  dependent traps}}.
\newblock {\emph{\JournalTitle{Phys. Rev. Lett.}}}
  \textbf{\bibinfo{volume}{77}}, \bibinfo{pages}{5315--5319},
  \doiprefix\url{https://doi.org/10.1103/PhysRevLett.77.5315}
  (\bibinfo{year}{1996}).

\end{thebibliography}

\newpage
\setcounter{page}{1}

\renewcommand{\thefigure}{S\arabic{figure}}
\setcounter{figure}{0}
\renewcommand{\thetable}{S\arabic{table}}
\setcounter{table}{0}
\renewcommand{\thesection}{S\arabic{section}}
\setcounter{section}{0}
\renewcommand{\thesubsection}{\Alph{subsection}}
\renewcommand{\theequation}{S\arabic{equation}}
\setcounter{equation}{0}
\renewcommand{\thepage}{S\arabic{page}}
\setcounter{page}{1}
\renewcommand{\appendixname}{}
\counterwithout*{equation}{section}	
\setcounter{footnote}{0}
\rfoot{\small\sffamily\bfseries\thepage}%

\begin{center}
\Large{\bf Supplementary information to \\\emph{On the survival of the quantum depletion of a condensate after release from a magnetic trap}}
\end{center}

\noindent\large{J. A. Ross, P. Deuar, D. K. Shin, K. F. Thomas, B. M. Henson, S. S. Hodgman,  and A. G. Truscott*}\\
\normalsize
* Corresponding author email: andrew.truscott@anu.edu.au

\section*{Experimental details}

\subsubsection*{Peak density calibration}

    The quantum depletion and contact are both predicted to depend solely on the condensed number and trapping frequencies via the condensate density, hence it is important to determine both quantities accurately. 

    The sole experimental parameters in the expression for the peak density (Eqn. (11) )
  are the geometric trap frequency $\bar{\omega} = \left(\omega_x\cdot\omega_y\cdot\omega_z\right)^{1/3}$, and $N_0$, the number of atoms in the condensate. 
    We simultaneously determine the total atom number $N$ and trap frequency $\bar{\omega}$ in a single shot using a pulsed atom laser and use the thermal fraction $\eta_T$ (see below) to determine the condensed number $N_0 = (1-\eta_T)N$. 

	The pulsed atom laser consists of a series of Fourier-broadened RF pulses centred on the minimum Zeeman splitting in the trap. 
	The pulse transfers atoms in the trap to the untrapped $m_J=0$ state with an approximately constant transfer rate across the cloud \cite{Manning10,Henson18}. 
	We outcouple approximately 2\% of the atoms per 100$~\mu$s pulse for $\approx$200 pulses, which eventually depletes the entire trap. 
	The atom laser thus prevents the detector from saturating and allows an accurate determination of the atom number, up to a factor of the quantum efficiency. 
	We determine the trapping frequencies by inducing centre-of-mass oscillations with a magnetic impulse, and find the oscillation period from the atom laser pulses \cite{henson18ML}.

\subsubsection*{Determining spin transfer efficiency}
\label{sec:th_spin}

\begin{figure}[h!]
	\begin{center}
		\includegraphics[width=0.8\textwidth]{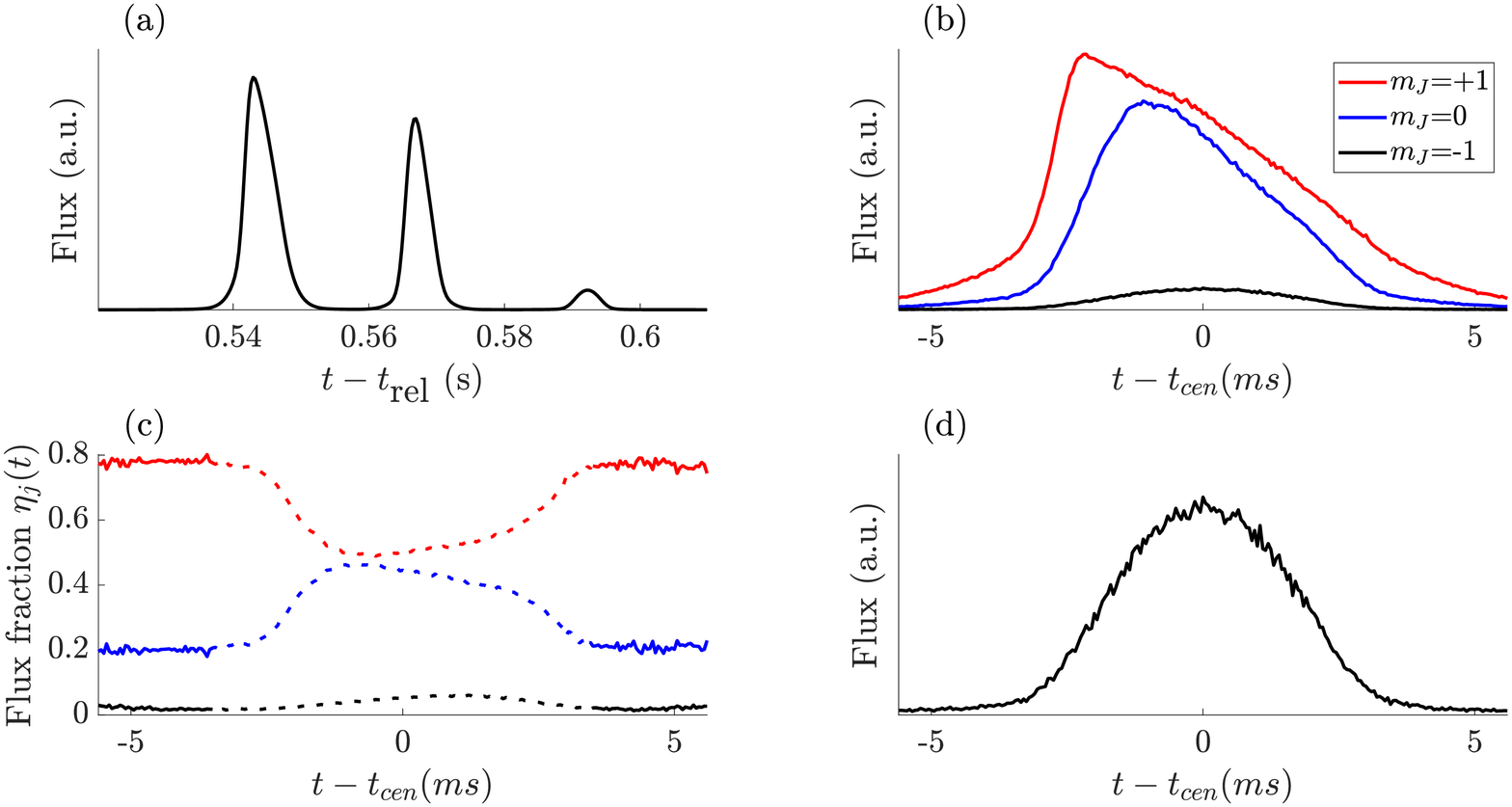}
		\caption{Determining the RF transfer efficiency. The time-of-flight profiles of each pulse are resolved (a) by applying a weak Stern-Gerlach pulse during the time of flight. The pulses are aligned with respect to their centre-of-mass (b) and used to determine the pointwise fraction ((c), dotted line). Detector saturation is evident in the peaks (dashed lines), but not in the thermal tails (solid lines), which are used to compute the transfer efficiency. Because of its lower flux, the $m_J=-1$ pulse does not show any clear evidence of saturation (d) and is used to determine the thermal fraction and hence $N_0$.}
		\label{fig:frac_cal}
	\end{center}
	\end{figure}

	To calibrate the transfer efficiencies, we applied a weaker Stern-Gerlach than for the depletion measurement, such that each $m_J$ cloud hits detector at different times (and positions), as illustrated in 
	Figure \ref{fig:frac_cal}. 
	The efficiencies $\eta_J$ cannot be calculated by counting the atoms in each cloud because the detector saturates during the peak condensate flux. However, we can compare the thermal parts. We align each cloud along the time (Z) axis and compute the pointwise fraction of the atomic flux $\phi(t)$ accounted for by each cloud, $\eta_j(t) = \phi_j(t)/\sum_j\phi_j(t)$, as depicted in 
	Figure \ref{fig:frac_cal} (a-c).
	The ratio of densities between the clouds is roughly constant in the thermal part (Figure \ref{fig:frac_cal} (c)), indicating the absence of saturation effects in the thermal part and a spin transfer efficiency that is independent of $k$. 
	The fraction of the original cloud transferred into each $m_J$ state is determined by taking the average $\langle\eta_j(t)\rangle$ over the thermal tails.

	We find these efficiencies are approximately 74\%, 24\%, and 2\% in all runs for the $m_J=+1$, 0, and -1 states, respectively.
	While the $m_J=0$ and $m_J=1$ clouds clearly saturate the detector, the small fraction ($\approx2\%$) of the atoms transferred to the $m_J=-1$ state does not (Figure \ref{fig:frac_cal} (d)). 
	A bimodal fit to the condensed and thermal parts, plus constant background, yields the thermal fraction $\eta_T$ and condensed fraction $1-\eta_T$.

\subsubsection*{Noise sources}
\label{sec:spinpop}

	In early tests of our measurement sequence we noticed a contamination of the signal by spurious counts. 
	Specifically, an extremely dilute remnant of the $m_J=+1$ cloud was observed as a small peak in counts in the region $-10\micron^{-1}\lesssim k_z \lesssim -8.5\micron^{-1}$.
	We inferred these were remnant counts from the $m_J=+1$ cloud as they were still visible when we ran an experimental sequence without the Landau-Zener transfer, but including the Stern-Gerlach pulse.

	While the cause of the appearance of the remnant counts is unclear, we observe that the count density outside the region of interest is similar in both the shots with the RF pulse (Landau-Zener transfer) and those without. 
	We note that only about one in a million atoms from the $m_J=1$ cloud are present in this manner in a given shot.
	On average, this accounted for approximately $N_{mJ=+1}=1$ atom per shot

	when running the sequence without the Landau-Zener transfer, and approximately half this amplitude after applying the pulse. 
	As the signal of interest was not the angular density profile but the integrated number of counts, we corrected for this contribution by subtraction, $N_{k_{\text{min}},k_{\text{max}}} 
	= (N_{k_{\text{min}},k_{\text{max}}}^{\text{detected}}-\eta_1 N_{m_J=+1})/(\eta_0\times\eta_Q\times\Omega_\text{ROI})$, where $\Omega_\text{ROI}$ is the fraction of the sphere retained by the windowing procedure described in the main text. and $\eta_Q$ is the detector quantum efficiency. 
	Such counts constitute about 10(5)\% of the detection events in the ROI and are not sufficient to account for the excess tail strength.
	We hypothesize that the remnant counts are atoms transferred into the $m_J=0$ state by non-ideal behaviour of the Stern-Gerlach pulses or magnetic field switches.

\subsubsection*{{Dependence of tail properties on the ROI}}

Table \ref{tab:choice_indep} shows the absolute and relative tail amplitude estimates $\Lambda_\textrm{fit}$ and $\Lambda_\textrm{fit}/\Lambda_\textrm{pred}$, respectively, that proceed from variations of the ROI and assumed quantum efficiency in the analysis of the experimental data. It is seen that
{neither the uncertainty in the collection efficiency nor the choice of elevation angle cutoff $\phi_c$ have a significant effect on the findings.}

	\begin{table}[h!]
	\centering
		\begin{tabular}{c c c c}
			\hline\hline

			QE & fit $r^2$ &  $\Lambda_\textrm{fit}$ & $\Lambda_\textrm{fit}/\Lambda_\textrm{pred}$\\      
			\hline
			0.05    &   0.83   &   0.2(0.1,0.3)  &  6.8(4.5,9.1)\\
			0.06    &   0.83   &   0.3(0.2,0.4)  &  7.4(4.9,9.8)\\
			0.07    &   0.83   &   0.3(0.2,0.4)  &  7.8(5.2,10.5)\\
			0.08    &   0.83   &   0.4(0.3,0.5)  &  8.3(5.5,11)\\
			0.09    &   0.83   &   0.5(0.3,0.6)  &  8.7(5.8,11.6)\\
			0.1     &   0.83   &   0.6(0.4,0.7)  &  9.0(6.0,12.1)\\
			0.11    &   0.83   &   0.6(0.4,0.8)  &  9.4(6.2,12.5)\\
			\hline
			$\phi_c$ & fit $r^2$ &  $\Lambda_\textrm{fit}$ & $\Lambda_\textrm{fit}/\Lambda_\textrm{pred}$\\
			\hline
			80$^\circ$    &   0.82   &   0.1(0.0,0.1) &  9.3(6.0,12.7)\\
			70$^\circ$    &   0.86   &   0.2(0.1,0.3) &  9.2(6.4,12)\\
			60$^\circ$    &   0.83   &   0.4(0.3,0.5) &  8.3(5.5,11)\\
			\hline\hline
		\end{tabular}
		\caption{{Tail predictions with modified ROI or assumptions about the value of quantum efficiency (QE).}
A change in quantum efficiency (QE) presents by changing the value of $\epsilon$ used in the prediction and also through the factor of $N_{0}^{7/5}$ used to compute the condensate density $n_0$. These effects partially cancel to produce a weak scaling of $\Lambda_\textrm{fit}/\Lambda_\textrm{pred}$ with respect to $\epsilon$.  Re-running the analysis using different QE {and the standard ROI with $\phi_c=\pi/3$}  yields fits that barely differ in the goodness-of-fit criterion and present comparable results for $\Lambda_\textrm{fit}$. We also find that the choice of collection area defined by the elevation angle cutoff $\phi_c$ {(while using the best QE value of 0.08)} has a weak effect on the result, but below statistical significance. Terms in brackets are the upper and lower 95\% confidence intervals.
		Neither the uncertainty in the collection efficiency nor the choice of elevation angle cutoff $\phi_c$ have a significant effect on the findings. }
		\label{tab:choice_indep}
	\end{table}

\section*{Statistical details}

\begin{figure}[h!]
	\centering
	        \includegraphics[width=0.8\textwidth]{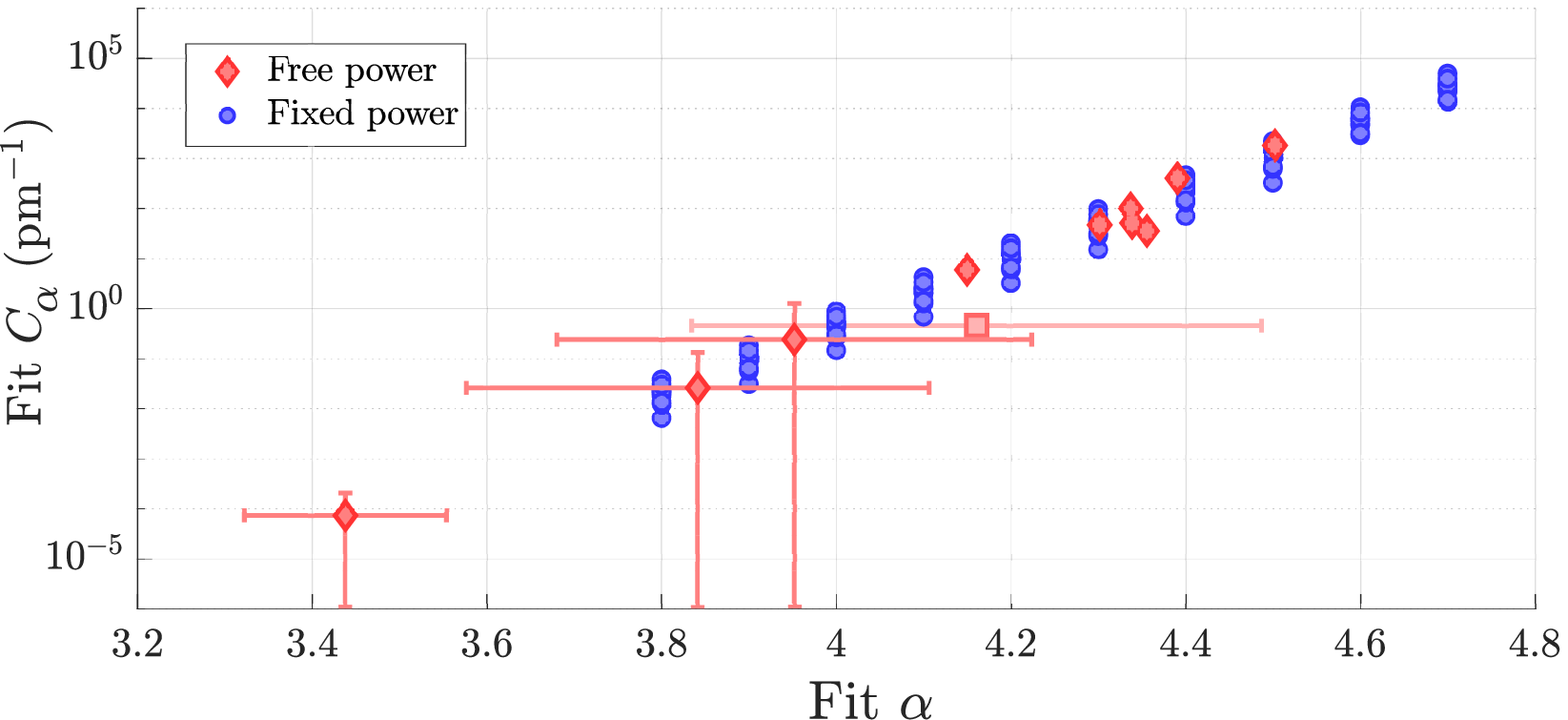}
	        \caption{Illustrating the large systematic errors in a fit to the density using a power law ansatz. If the fit to the density profile {(Fig \ref{fig:nk_density} in the main text)} ) 
	        is replaced by Eqn. (\ref{nkfit}) (also in main)	 
	        and $\alpha$ {and $C_{\alpha}$} used as 
	        fit parameter{s}, it gives a wide variation in best-fit exponents and scale coefficients between each data set (red diamonds). The red square shows the mean fitted $\alpha$ and (geometric) mean $C_\alpha$ with standard deviation (in $\alpha$) shown as error bars. Blue circles show the amplitude coefficient $C_\alpha$ obtained from the same data sets with $\alpha$ fixed at {a number of} values within the range of the free fit results. The choice of $\alpha$ strongly determines the coefficient $C_\alpha$, but the error bars (standard errors in fit parameters) are smaller than the markers in all cases. }
	        \label{fig:contact_determination_issues}
	\end{figure}

    In Figure \ref{fig:contact_determination_issues} 
	we illustrate how fixing the scaling exponent $\alpha$ when fitting to a given dataset obscures the enormous covariance between $\alpha$ and the scale coefficient $C_\alpha$.
	The red diamonds in  
	Figure \ref{fig:contact_determination_issues} 
	show the fit exponent and amplitude obtained from a fit {of Eqn. (\ref{nkfit})} 
	to {single data sets}, 
	each representing a different BEC density. 
	The variation in the fit exponent is small (mean 4.2, standard deviation 0.4), and although the mean $\alpha$ from the fits is only half a standard deviation (one standard-error interval) different from 4, the corresponding difference in amplitude $C$ varies exponentially with $\alpha$, spanning over six orders of magnitude.    
	
	The blue circles in 
	Figure \ref{fig:contact_determination_issues}
	show the amplitudes returned from fits to data from all datasets when $\alpha$ is constrained to the set values as shown. These set values are chosen 
	within about one standard deviation of the mean $\alpha$ from the two-parameter fits to separate datasets. 
	These blue fits scarcely differ in their goodness-of-fit criterion (the mean square error) and so offer no obvious way to reconcile the expected distribution with these divergent statistical conclusions.
	Furthermore, this is not reflected in the error estimates in the fitting routines: The error bars representing the uncertainty in parameters from the fixed-$\alpha$ fits are smaller than the markers used in 
	Figure \ref{fig:contact_determination_issues}. 
	A linear fit reveals that $d \log_{10} C_\alpha/d\alpha \approx 6.8$.

	Ultimately, the question of whether the data {is drawn from a power law at all is not amenable to a decisive conclusion. 
	For example, a log-normal distribution can produce similarly accurate predictions to the power law (see Figure \ref{fig:exp_results} in the article) although there is no physical hypothesis that predicts such a distribution. }
	This points to one of the primary challenges with power laws; the exponents are strongly entwined with the rate of occurrence of rare events, which by definition are subject to large statistical fluctuations and and thus subvert even the most meticulous investigations.
	These problems with fitting power laws are ubiquitous, and made more difficult by the small range of $k$ which are visible in the helium experiments.
	In general, estimating the exponent of a purported power law is difficult and requires data spanning several orders of magnitude in scale \cite{Goldstein04,Clauset09,Virkar14,Hanel17}, which are not present in either {of the} helium experiment{s to date}.
	
Nevertheless  --- the analysis of Fig.~\ref{fig:contact_determination_issues} does allow us to at least say that the data is inconsistent with power laws outside the range $3.3<\alpha<4.7$.

\section*{Simulations}
    
\subsubsection*{{Observables}}

Detailed data extracted from the simulations are given in Table.~\ref{tab:simdata}.

	The total quantum depletion of the condensate  $\delta_B$ is given by
	\eq{depletion}{
	\delta_B  = \frac{N_B}{N}=\frac{1}{N}\int d^3\bo{x}\left\langle \real{\psi_B(\bo{x},t)\wt{\psi}^*_B(\bo{x},t)} \right\rangle_{\rm stoch.}
	}
	 Stochastic averaging over all trajectories in the ensemble is denoted by $\langle\cdot\rangle_{\rm stoch.}$.
	The density of the depleted particles is evaluated by the standard positive-P expression, 
	\eq{densityB}{
	n_B(\bo{k}) = \left\langle\dagop{\Psi}_B(\bo{k},t)\op{\Psi}_B(\bo{k},t)\right\rangle 
	= {\rm Re}\left\langle \wt{\psi}^*_B(\bo{k},t)\psi_B(\bo{k},t) \right\rangle_{\rm stoch.}.
	}
	The density of condensate
	\eq{densityC}{
	n_{BEC}(\bo{k}) = \left(1-\delta_B\right)|\phi(\bo{k},t)|^2,
	}
	is augmented by the $1-\delta_B$ factor when calculating observables to conserve overall particle number. 
	The k-space fields {here are normalised as}
	\eq{kspace}{
	\matri{c}{\op{\Psi}_B(\bo{k})\\\psi_B(\bo{k})\\\wt{\psi}_B(\bo{k})} = \frac{1}{{(2\pi)^3}
	}\int d^3\bo{x}\ e^{-i\bo{k}\bo{x}}\ \matri{c}{\op{\Psi}_B(\bo{x})\\\psi_B(\bo{x})\\\wt{\psi}_B(\bo{x})}
	}

	\begin{figure}[h!]
		\begin{center}
		\includegraphics[width=0.49\textwidth]{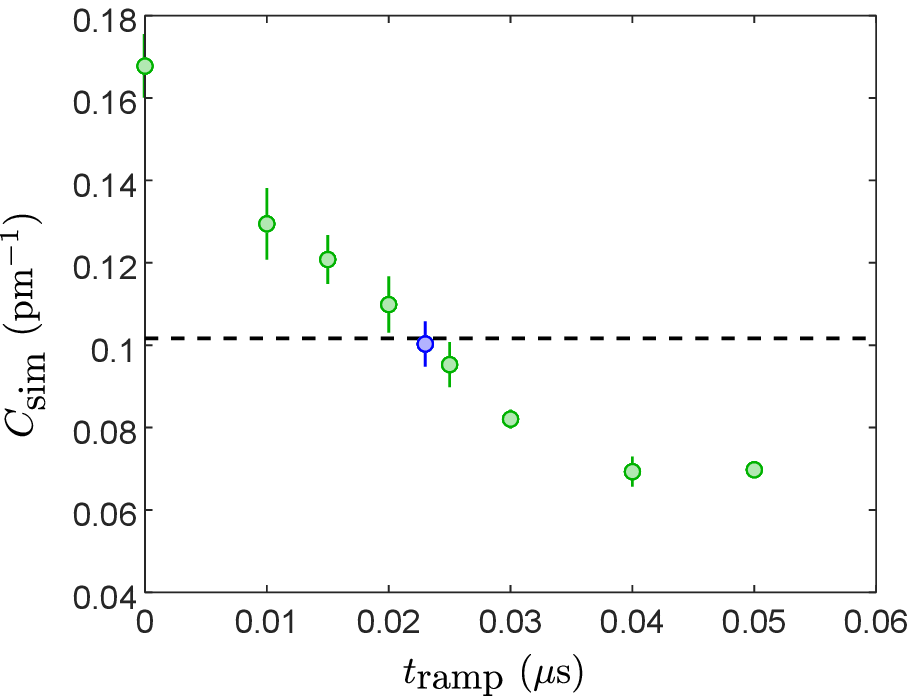}
		\includegraphics[width=0.49\textwidth]{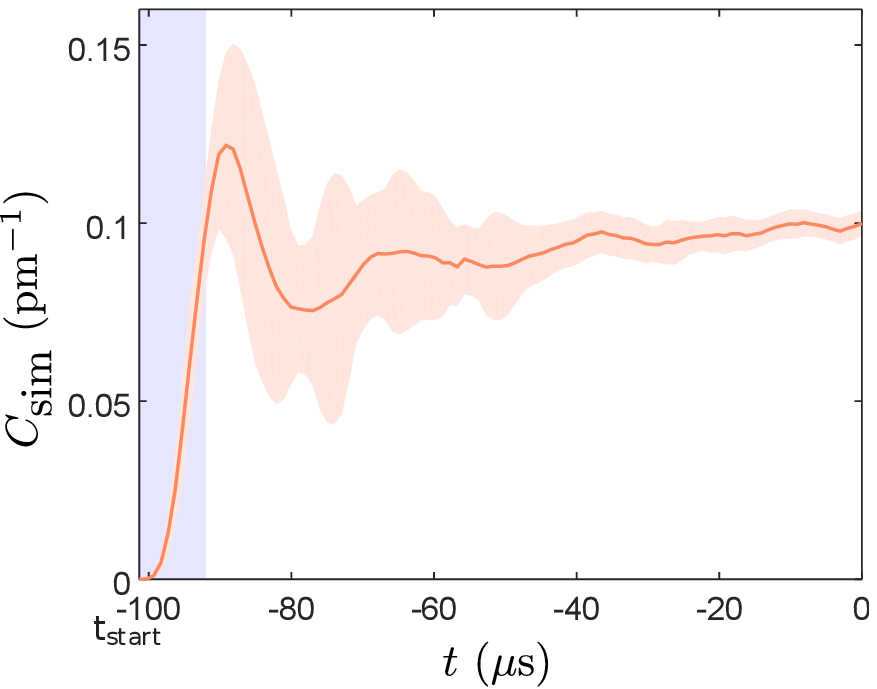}
		\end{center}\vspace*{-0.5cm}
		\caption{
		Details of initial state preparation, shown on the $N=455852$ case.
		Upper panel: 
		Calibration data. Shown is the calculated contact $C_{\rm sim}$ at the end of the initial state generation ($t=0$), as a function of the quench ramp time $t_{\rm ramp}$. 
		The horizontal dashed line shows the LDA prediction $C=0.102$ pm$^{-1}$, the blue data point the ensemble that was deemed to agree, and was used for subsequent simulations for $t>0$.
		Lower panel: 
		Evolution and stabilisation of the contact. 
		The blue shaded area shows the duration  of the ramp from $t_{\rm start}$ to $t_{\rm start}+t_{\rm ramp}$. Orange shading denotes the error bars.
		$\mc{S}=4000$ trajectories in all cases. 
		\label{fig:init}}
	\end{figure}

\subsubsection*{{Initial ensemble}}

	A procedure for generating the equilibrium state has been developed in the Wigner representation \cite{Sinatra00,FINESS-Book-Ruostekoski,FINESS-Book-Sinatra,Martin10a,Martin10b}.
	Unfortunately we cannot use this directly, nor a direct Wigner representation of the quasiparticles because the condition required for correctness of the Wigner representation -- that there are more particles than modes -- is very far from being met \cite{Sinatra02,Norrie06,Deuar11} (In fact here we have about $\mc{O}(1000)$ particles in the depletion, and $\mc{O}(10^6)$ modes.). 
	It is also unclear how to translate a local density formulation of depletion in a uniform section of gas to a positive-P ensemble without introducing discontinuities. Instead we turned to dynamically generating a state with the required quantum depletion.

	We begin with a fully condensed ground state with $\psi_B=\wt{\psi}_B=0$ and $\phi(\bo{x})=\phi_0(\bo{x})$, the ground state of the Gross-Pitaevskii equation (\ref{GPE-eq}).
	The latter is obtained by imaginary time propagation of the GPE augmented with an appropriately chosen chemical potential $\mu$ according to (\ref{mu})
	which sets the central density.
	
	Our first attempt to generate the equilibrated quantum depleted state thus started with $\phi_0(\bo{x})$ and then adiabatically ramped the interaction from $g=0$ to the experimental value while evolving the equations \eqn{GPE-eq}--\eqn{pstab-eq}. 
	This did not work for two reasons. 
	Firstly, a very strong collective oscillation was induced, since the width of the Thomas-Fermi ground state depends strongly on $g$. Secondly, very long time evolutions succumb to excessive noise in the positive-P simulation and produce a state that is too noisy to be useful. Note that in the positive-P representation different ensembles can represent the same state but exhibit very different noisiness and practical usefulness \cite{DeuarPhD,Drummond20}.

	The second attempt began with the Gross-Pitaevskii ground state and the {target} physical interaction strength $g$ in the GPE \eqn{GPE-eq},
	while slowly ramping the interaction strength in the Bogoliubov equations \eqn{pstab-eq}. 
	This eliminates the main oscillations in the mean-field evolution. The least unwanted nonadiabatic disturbance occurs when $g$ is ramped only within the projected part of the Bogoliubov evolution as per
	\eqs{nstab-ramp}{
	i\hbar\frac{d\phi}{dt} &=& \mc{H}(g,\phi)\phi = \eqn{GPE-eq}\\ 
	i\hbar\frac{d\psi_B}{dt} &=& \mc{H}(g,\phi)\psi_B 
	+\mc{P}_{\perp}\left\{
	g_B|\phi|^2\psi_B + g_B\phi^2\wt{\psi}_B^* + \sqrt{-ig_B}\,\phi\,\xi(\bo{x},t)
	\right\}\\
	i\hbar\frac{d\wt{\psi}_B}{dt} &=& \mc{H}(g,\phi)\wt{\psi}_B
	+\mc{P}_{\perp}\left\{
	g_B|\phi|^2\wt{\psi}_B + g_B\phi^2\psi_B^* + \sqrt{-ig_B}\,\phi\,\wt{\xi}(\bo{x},t)
	\right\}.
	}
	with $g_B(t)$.
	However, over adiabatic timescales, this still introduced far too much noise in the Bogoliubov fields $\psi_B$, $\wt{\psi}_B$  to be useful.

	To work around the problem, we take advantage of the fact that an instantaneous quantum quench of a weakly interacting condensate produces depletion with a similar time-integrated momentum profile $\int dt\, n_B(\bo{k},t) \propto 1/k^4$ to the ground state value, but with a somewhat higher depletion, as described in detail in \cite{stob,nstab-longpaper}. Simulations show that non-instantaneous but rapid quenches produce lower amounts of quantum depletion, as shown in 
	Figure~\ref{fig:init} (left panel). 
	To generate these ensembles, we used the following ramp:
	\eq{gramp}{
	g_B(t) = \ifmath{c@{\quad\text{when}\quad}l}{
	\left(\frac{t-t{\rm start}}{t_{\rm ramp}}\right)\,g & t<t_{\rm start}+t_{\rm ramp}\\
	g & t\ge t_{\rm start}+t_{\rm ramp}
	}
	}
	starting at negative $t_{\rm start}$, and evolving till $t=0$.  Thus $t_{\rm ramp}$ was the ramp time, and the remaining time: $|t_{\rm start}|-t_{\rm ramp}$ an equilibration time. 
	We carried out a calibration like that shown in 
	Figure~\ref{fig:init} ({left} 
	panel) for each set of cloud and trap parameters needed. Then for the initial \textit{in situ} ensemble in the trap we chose the ensemble generated with the $t_{\rm ramp}$ that produces total quantum depletion in agreement with the \emph{in situ} value (\ref{eqn:TotalHarmonicContact}) 
	for the condensate ground state (like the blue data point in 
	Figure~\ref{fig:init}, 
	{left} panel).
	The value of $t_{\rm start}$ was $-101\mu$s for the weakly trapped cases ($\omega=45\times425\times425$ Hz and $\bar{\omega}=201$ Hz) and $-198\mu$s for the strongly trapped cases ($\omega=71\times902\times895$ Hz and $\bar{\omega}=393$ Hz) detailed in Table.~\ref{tab:simdata}.
	Figure~\ref{fig:init} ({right} 
	panel) shows an example of the evolution of the contact \textit{in situ} during this initial state generation.
	Properties of the initial states are labelled (CT) and (ST) in Fig. \ref{fig:sim_fig} (a) (main text).

\subsubsection*{Determination of the contact {and tail strengths}}
	\label{TH-TAIL}

	Much as in the experiment, the k-space density $n(\bo{k})$ in the simulations is very noisy in the asymptotic region of large $|\bo{k}|$, and a lot of averaging is needed to extract the contact. 
    {In light of the discussion on power laws, the exponent $\alpha$ was not fitted here, but we did calculate both the fit coefficients $C_{\rm sim}$ on the function $n_B(k)=C_{\rm sim}/k^4$ with $\alpha=4$ assumed and the count of particles in the tails, $N_{k_{\rm min},k_{\rm max}}$ as in Table~\ref{tab:Nminmax}, in order to have a systematic view of how correlated the two kinds of results are.}

	We proceed as follows:
	The simulations provide a density of depleted particles $n_B(\bo{k})$. 
	Like with the experimental counts, we keep only density that is far enough away {($\phi_c$, usually $60^o$)} from the long axis of the cloud. 
	The simulated system is axially symmetric around the long axis. 
	We do not restrict counting to the vicinity of the vertical axis because here there are no detector irregularities to avoid, and this allows us to improve the signal to noise ratio. 
	
\begin{figure}[h!]
	\begin{center}
	\includegraphics[width=0.5\textwidth]{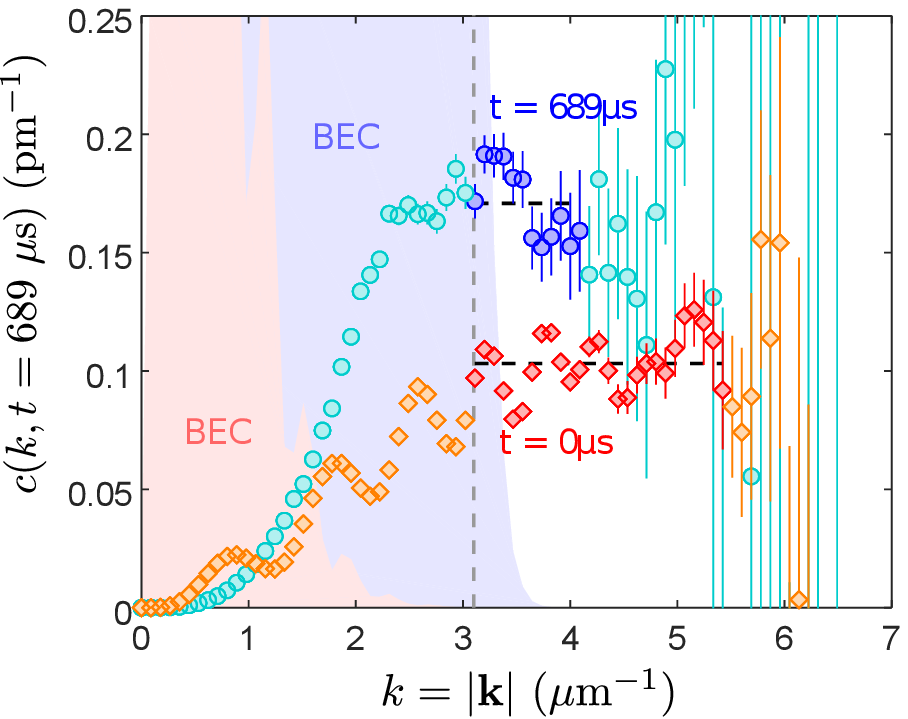}
	\end{center}\vspace*{-0.5cm}
	\caption{
	The extraction of contact from simulation data -- shown by example of the $N=455852$ trap release (CE) simulation. 
	Blue/cyan data: after expansion at $t=689\mu$s, Red/orange: in situ {initial condition} at $t=0$. 
	Each data point corresponds to estimation of $c(k)$ given by \eqn{ck} using the density of particles in the Bogoliubov field $\op{\Psi}_B(\bo{k})$ in a radial bin centred at $k=|\bo{k}|$. 
	Error bars give the statistical error from an ensemble of $\mc{S}=4000$ trajectories. 
	Blue and red points are the fitted part of the data. 
	The vertical dashed line indicates $k_{\rm inner}$, while the shaded region on the left shows the corresponding scaled density {$c_{\rm BEC}(k)=n_{\rm BEC}(k)k^4$} 
	that would be calculated from the condensate field $\phi(\bo{k})$.
	The horizontal dashed lines indicate the final estimates of $C_\textrm{sim}$ extracted from the data.
	}
	\label{fig:Ck}
	\end{figure}

	The particle counts $N_B(\bo{k})=n_B(\bo{k}){/V}$ 
	are finely binned according to the absolute value of the momentum $k=|\bo{k}|$, giving total bin count $N_k$. 
	The field is simulated on a square lattice in k space, in which each site corresponds to a k-space volume of $\Delta V_K{=(2\pi)^3/V}$ {with real space volume $V$}. Therefore the total bin volume $V_k$ is obtained by binning the site volumes, and the mean density in each bin is {$N_k/V_k$ so $n_k=(2\pi)^3N_k/V_k$}. Each bin gives an estimate of the corresponding apparent contact: 
	\eq{ck}{
	c(k) = {n_k k^4} = \frac{N_k}{V_k}\,k^4 (2\pi)^3.
	}
	The statistical error estimate on $c(k)$, $\Delta c(k)$, is obtained by averaging subensembles, then using the central limit theorem (CLT) on the subensemble averages. 
	Figure~\ref{fig:Ck} shows example values and error estimates{, while also demonstrating the difficulties involved.}

	We can see that for a significant range of $k$ values the $c(k)$ estimate gives fluctuations around a constant value. 
	However, several difficulties in extracting the overall trend are also evident. Low $k$ values are not representative because the particles in the shaded region never emerge from the condensate and are not measured in the experiment.
	Therefore we remove data below a value $k_{\rm inner}$ from consideration. 
	$k_{\rm inner}$ is chosen such that it does not include any depletion that would significantly overlap with the BEC in the final expanded cloud and be obscured by it. 
	
	This corresponds also to the energy at which the quasiparticle spectrum becomes particle-like, since the mean field energy in the centre of the trapped cloud is responsible for both effects.

	High $k$ values, on the other hand, suffer from statistical error sufficiently high as to make them useless, and so we also choose a maximum  $k_{\rm outer}$ value, and only use $k\in[k_{\rm inner},k_{\rm outer}]$ to extract the contact estimate. 
	The {extent of the} useless high k region changes with time during the simulation. To systematically adapt our fitting region to this, we choose $k_{\rm outer}$ according to the calculated statistical error such that 
	\eq{kouter}{
	\Delta c(k,t) \le \Delta_{\rm max}\quad \forall k\le k_{\rm outer}(t), 
	}
	and $\Delta_{\rm max}$ is chosen once for all times in a simulation. This gives a time-dependent fitting region $k\in[k_{\rm inner},k_{\rm outer}(t)]$.
	The final contact estimate $C_\textrm{sim}(t)$ is the mean of all $c(k,t)$ values in the fitting range. 
	Error bars on $C_\textrm{sim}(t)$, $\Delta C_\textrm{sim}(t)$, are obtained from a CLT estimate of the error in the mean of the points used, after binning enough neighbouring points to encompass the autocorrelation (i.e. so that neighbouring bins are uncorrelated). 
	The $\Delta_{\rm max}$ is chosen low enough so that the statistical error in individual points $\Delta c(k,t)$ at the large $k$ end does not put off the overall contact estimation $C_\textrm{sim}(t)\pm\Delta C_\textrm{sim}(t)$. 
	Operationally we choose the value of $\Delta_{\rm max}$ at which the error introduced by adding successive high $k$ points introduces more uncertainty than the reduction thanks to a larger ensemble.

\begin{table}[h!]
{
	\begin{center}
	\setlength{\tabcolsep}{0.5pt}
	{\fontsize{8}{11}\selectfont
	\begin{tabular}{c|ccc|c|c}
	\hline\hline
	$n_oN$	&$k_{\rm min}$&$k_{\rm max}$ & $\phi_c$ & $N_{k_{\rm min},k_{\rm max}}$	&ratio $\mathcal{A}_{\rm sim}$ 	\\
	($10^6\mu$m${}^{-3}$)& \multicolumn{2}{c|}{($\mu$m${}^{-1}$)}& (deg) &final& 	final/in situ\\
	\hline
	\multicolumn{6}{l}{Rapid release from trap (CE)}\\
	5.237(16)   & 2.0 & 3.5 & 60 & 193(8)    & 1.38(5)\\
	12.09(6)    & 2.25 & 3.5 & 60 & 362(10)   & 1.43(4)\\
    6.86(3)     & 2.75 & 4.0 & 60 & 163(4)   & 1.45(4)\\
    19.90(8)    & 3.0 & 4.0 & \ \ 40 & \ \ 423(8)    & \ \ \ 1.37(3) \\
    19.90(8)    & 3.0 & 4.0 & \ \ 50 & \ \ 408(8)    & \ \ \ 1.56(3) \\
    19.90(8)    & 3.0 & 4.0 & 60 & 368(7)    & 1.80(4) \\
    19.90(8)    & 3.0 & 4.0 & \ \ 70 & \ \ 282(6)    & \ \ \ 2.00(4) \\
    19.90(8)    & 3.0 & 4.0 & \ \ 80 & \ \ 157(5)    & \ \ \ 2.19(6) \\
	\hline
	\multicolumn{6}{l}{Rapid release, spherical (SE)}\\
	5.35(2)     & 2.0 & 3.25 & 60 & 120(9)    & 0.92(6)\\
	12.23(6)    & 2.0 & 3.0 & 60 & 256(11)   & 1.05(4)\\
    20.38(10)   & 3.0 & 4.0 & 60 & 232(10)    & 1.10(5) \\
	\hline
	\multicolumn{6}{l}{Slow ramp down of trap (CS)}\\
	5.237(16)   & 2.0 & 3.5 & 60 & 55(10)   & 0.37(6)\\
	12.09(6)    & 2.0 & 3.5 & 60 & 103(21)   & 0.31(5)\\
	6.86(3)     & 2.5 & 4.0 & 60 & 50(8)   & 0.34(6)\\   
	19.90(0)    & 2.5 & 4.0 & 60 & 181(14)   & 0.54(4)\\
	\hline\hline
	\end{tabular}
	}
	\end{center}
	\caption{
	Tail strength data in simulations. Based on final times in the simulations described in Table~\ref{tab:simdata}, referenced by the value of $n_0N$. $N_{k_{\rm min},k_{\rm max}}$ is calculated as per Eqn (\ref{eqn:pred_num}) in the main text. 
	The ratio $\mathcal{A}_{\rm sim}$ of tail strength  in the expanded cloud compared to Tan theory predictions is obtained by dividing $N_{k_{\rm min},k_{\rm max}}$  at the final time by its value in situ at $t=0$. $k_{\rm max}$ is chosen to still contain all $4\pi$ steradians inside the square lattice, while  $k_{\rm min}$ to avoid overlap with the expanding condensate.
	\label{tab:Nminmax}}
	}
\end{table}

	We also found that the relative factors between Eqn. (\ref{eqn:pred_num}) 
	and the simulated tails (i.e. $C_\textrm{sim}/\mathcal{C}$ {and $\mathcal{A}_{\rm sim}$) depend} on choice of cutoff angle $\phi_c$. 
	{We find that apparent tail strengths $C_\textrm{sim}$ {and $\mathcal{A}_\textrm{sim}$ are} larger for smaller collection regions that are more tightly concentrated about the  
	strong trapping axis, whereas larger collection angles (that include areas closer to the weak axis) produce lower apparent tail strengths. 
	Data on $\mathcal{A}_{\rm sim}$ are shown in Table~\ref{tab:Nminmax} while 
	$C_\textrm{sim}$ was 1.6(1), 1.9(2), and 2.2(3) times $\mathcal{C}$ for $\phi_c$ values of 60, 70, and 80 degrees, respectively.}

	\begin{table*}[h!]
	\begin{center}
	\setlength{\tabcolsep}{0.5pt}
	{\fontsize{8}{11}\selectfont
	\begin{tabular}{cccccc|cc|ccc}
	\hline\hline
	Trap at $t=0$	&Peak density $n_0$ 	& $N$	& $n_oN$	&timescale	& Time	$t$	& $C_\textrm{sim}(t)$			& $C_\textrm{sim}(t)/C_\textrm{sim}(0)$	& $N_B(t)$	& $k_{\rm inner}$--$k_{\rm outer}$	& $\Delta_{\rm max}$	\\
	$\omega$ (Hz)	&at $t=0$ ($\mu$m${}^{-3}$)	&& ($10^6\mu$m${}^{-3}$)&  ($\mu$s)&  ($\mu$s)& (pm${}^{-1}$)	&		&		& ($\mu$m${}^{-1}$)			& ($\mu$m${}^{-1}$)	\\
	\hline
	\multicolumn{6}{l|}{In situ initial state (CT,ST):}&&&&&\\
	$425\times425\times45$&16.82(5)	&311360&5.237(16)	&$t_B$=19.8		&0		&0.0260(13)		& --		&1391(7)	&2.1--4.2				&0.025			\\	
	$425\times425\times45$&21.32(10)	&567180&12.09(6)	&$t_B$=15.8		&0		&0.060(5)		& --		&2860(13)	&2.2--4.75				&0.032			\\	
	$902\times895\times71$&32.17(13)	&213293&6.86(3)	&$t_B$=8.5		&0		&0.034(4)		& --		&1324(6)	&2.9--4.9				&0.020			\\	
	$902\times895\times71$&43.66(18)	&455852&19.90(8)	&$t_B$=9.3		&0		&0.100(3)		& --		&3047(12)	&3.25--5.4				&0.032			\\	
	$201\times201\times201$&16.88(6)	&316766&5.35(2)	&$t_B$=19.8		&0		&0.025(3)		& --		&1396(7)	&2.1--4.1				&0.025			\\	
	$201\times201\times201$&21.32(10)	&573650&12.23(6)	&$t_B$=14.2		&0		&0.060(9)		& --		&2741(11)	&2.2--4.15				&0.04			\\	
	$393\times393\times393$&44.15(21)	&461514&20.38(10)	&$t_B$=7.7		&0		&0.101(5)		& --		&3107(12)	&3.0--5.0				&0.025			\\	
	\hline
	\multicolumn{6}{l|}{Rapid release from trap (CE,SE), $\tau_{\rm release}=37.5\mu$s:}&&&&&\\
	$425\times425\times45$&16.82(5)	&311360&5.237(16)	&$\tau_{\rm release}$&1583	&0.0386(22)		& 1.49(11)	&910(20)	&2.1--2.8				&0.020			\\	
	$425\times425\times45$&21.32(10)	&567180&12.09(6)	&$\tau_{\rm release}$&1346	&0.090(7)		& 1.51(17)	&1880(40)	&2.3--3.05				&0.020			\\	
	$902\times895\times71$&32.17(13)	&213293&6.86(3)	&$\tau_{\rm release}$&689	&{0.057(5)}		& {1.69(14)}	&950(20)	&2.8--3.7				&0.03			\\	
	$902\times895\times71$&43.66(18)	&455852&19.90(8)	&$\tau_{\rm release}$&689	&{0.171(10)}		& {1.66(12)}&2300(40)	&3.1--4.1				&0.06			\\	
	$201\times201\times201$&16.82(5)	&316766&5.35(2)	&$\tau_{\rm release}$&1900	&0.025(3)		& 1.02(15)	&780(30)	&1.9--2.65				&0.025			\\	
	$201\times201\times201$&21.32(10)	&573650&12.23(6)	&$\tau_{\rm release}$&1900	&0.066(7)		& 1.11(20)	&1630(50)	&2.3--3.15				&0.05			\\	
	$393\times393\times393$&44.15(21)	&461514&20.38(10)	&$\tau_{\rm release}$&973	&0.110(5)		& 1.09(8)	&1780(50)	&3.1--3.8				&0.05			\\	
	\hline
	\multicolumn{6}{l|}{Slow ramp down of trap (CS):}&&&&&\\
	$425\times425\times45$&16.82(5)	&311360&5.237(16)	&$t_{\rm ramp}$=2375	&2375		&0.010(3)		& 0.40(12)	&1750(140)	&1.9--3.45				&0.030			\\	
	$425\times425\times45$&21.32(10)	&567180&12.09(6)	&$t_{\rm ramp}$=2375	&2375		&0.020(3)		& 0.35(7)	&2700(300)	&2.0--3.9				&0.040			\\	
	$902\times895\times71$&32.17(13)	&213293&6.86(3)	&$t_{\rm ramp}$=1216	&1216		&0.011(6)		& 0.3(2)	&1520(150)	&2.5--4.2				&0.040			\\	
	$902\times895\times71$&43.66(18)	&455852&19.90(8)	&$t_{\rm ramp}$=1216	&1216		&0.044(5)		& 0.45(6)	&3800(400)	&2.5--4.1				&0.035			\\	
	\multicolumn{6}{l|}{Instantaneous switch-off of the trap:}&&&&&\\
	$902\times895\times71$&43.66(18)	&455852&19.90(8)	&$t_{\rm ramp}$=0	&689		&0.0187(6)		& 1.80(6)	&2380(40)	&3.1--4.05				&0.05			\\	
	\hline\hline
	\end{tabular}
	}
	\end{center}
	\caption{
	Main simulation data and parameters {for fitting of $k^{-4}$ tail amplitude $C_{\rm sim}$.}
	The time $t$ for which data are calculated is counted relative to the start of the trap release.
	Abbreviations (CT,CE,\dots) as in Figure \ref{fig:sim_fig}a of the main text.
	 The range $k_{\rm inner}$--$k_{\rm outer}$ {here} was chosen as explained in the text and used to obtain the estimate and uncertainty for $C_\textrm{sim}$.
	In all cases, $\mc{S}=4000$ trajectories averaged. $N_B$ is the number of Bogoliubov field particles as per \eqn{depletion}.
	}
	\label{tab:simdata}
	\end{table*}

\clearpage	
	
\section*{Toy two-mode model of escape}	
In a uniform gas in the Bogoliubov approximation, the $\op{a}_k$ mode is coupled only to $\dagop{a}_{-k}$ and the condensate. The Bogoliubov-de Gennes equations for these modes can then be written \cite{Castin98}
	\begin{equation}
	\frac{d}{dt}\matri{c}{\op{a}_{k}(t)\\\dagop{a}_{-k}(t)} = -\frac{i}{\hbar}
	\matri{cc}{\hbar^2k^2/2m +2gn(t)-\mu(t) & gn(t)\\-gn(t) & -\hbar^2k^2/2m-2gn(t)+\mu(t)} \matri{c}{\op{a}_{k}(t)\\\dagop{a}_{-k}(t)}.
	    \label{eq:toyBDG}
	\end{equation}
	Notice the provision for a time-dependent background condensate density $n(t)$.
	We stay in the real-particle basis rather than the Bogoliubov quasiparticles $\op{b}_k$ which allows to avoid calculating time-dependent changes of the coefficients $u_k$, $v_k$. The chemical potential is $\mu(t)=gn(t)$ \cite{Castin98}.
 
The equations \eqn{eq:toyBDG} can be used as input for equations of motion of the low order moments $\rho(k)=\langle\dagop{a}_{\pm k}\op{a}_{\pm k}\rangle$, 	$A(k)=\langle\op{a}_{k}\op{a}_{-k}\rangle$, and $A^*(k)=\langle\dagop{a}_{k}\dagop{a}_{-k}\rangle$, for example $d\rho(k)/dt = \langle (d\dagop{a}_k/dt)\op{a}_k\rangle+\langle\dagop{a}_k(d\op{a}_k/dt)\rangle$. Assuming equal initial occupations $\rho(k,0)=\rho(-k,0)$, one obtains an evolution equation for two coupled quantities $\rho(k,t)$ and $A(k,t)=A_r(k)+iA_i(k)$. It is
	\begin{equation}
	\frac{d}{dt}\matri{c}{\rho(k,t)\\A_r(k,t)\\A_i(k,t)} = 
	\frac{1}{\hbar}\matri{ccc}{0 & 0 & -2gn(t)\\0 & 0 & 2(\hbar^2k^2/2m-gn(t)) \\ -2gn(t) & -2(\hbar^2k^2/2m-gn(t)) & 0} \matri{c}{\rho(k,t)\\A_r(k,t)\\A_i(k,t)} + \matri{c}{0\\0\\-gn(t)}.
	    \label{eq:toy}
	\end{equation}
The initial conditions are $\rho(k,0)=v_k^2$, $A(k,0)= v_ku_k$, corresponding to the Bogoliubov ground state. Taking $n(t)$ constant one obtains the solution \eqn{eq:caricature}.

A more realistic model is obtained by estimating the $n(t)$ that particles in the $k$ mode experience as the trap is released and the condensate expands. 
Hydrodynamic expansion of a condensate with a Thomas-Fermi profile suddenly released from the trap \cite{Castin96} gives a self-similar expansion with widths approximated by $W_j(t) = W_j(0)\sqrt{1+(\omega_jt)^2}$. The initial widths being the Thomas Fermi radii in the trap $R_j=(1/\omega_j)\sqrt{2gn_0)/m}$, $j=\{x,y,z\}$. From this, and assuming $\omega_y=\omega_z=\omega_{\perp}=\bar{\omega}\lambda^{1/3}$ (approximately true in our setup) the condensate density at any point in the cloud decays as 
\eq{toy:decay}{
  n_c(r,t) = \frac{n_c(r,0)}{(1+\omega_{\perp}^2t^2)\sqrt{1+\omega_x^2t^2}},\qquad\text{where}\qquad
  n_c(r,0) = \left\{\arr{@{\ }l@{\text{\quad if }}l}{n_0\left[1-(r/R_{\perp})^2\right] & r<R_{\perp}\\0 & r\ge R_{\perp}}\right.
}
Escape is much easier, however, for atoms near the edge of the cloud than in the centre, because they can avoid reabsorption by faster free-flight from the high density region. We categorise the initial position of the studied section of the gas with the quantity
\eq{toy:R0}{
R_0 = \frac{\sqrt{y^2+z^2}}{R_{\perp}} = \frac{\omega_{\perp}\sqrt{y^2+z^2}}{\sqrt{2gn_0/m}}.
}
The initial density is then 
\eq{toy:n0}{
n(0) = \left\{\arr{@{\ }l@{\text{\quad if }}l}{n_0(1-R_0^2)& R_0\le1\\0 & R_0>1}\right.
} 
The motion of the atoms is estimated by assuming they travel in the outward direction at velocity $\hbar k/m$ relative to the hydrodynamic expansion (which also accelerates them to some degree). Hence their position is estimated as 
\eq{toy:rt}{
r(t) = R_0R_{\perp}\sqrt{1+\omega_{\perp}^2t^2}+\frac{\hbar kt}{m}. 
}
The condensate at this position comes from hydrodynamic expansion of the initial condensate at position \eq{toy:rc0}{
r_{c0}(t) = \frac{r(t)}{\sqrt{1+\omega_{\perp}^2t^2}} = R_0R_{\perp} + \frac{\hbar k t}{m\sqrt{1+\omega_{\perp}^2t^2}}.
}
The local density at position $r$ at time $t$ is then $n_c(r_{c0}(t),t)$ according to \eqn{toy:decay}.

Finally, to roughly include ramp time in the analysis, consider the density in a trapped gas that adiabatically follows the trap evolution $\omega_j(t)=\omega_j(0)e^{-t/\tau_{\rm release}}$. The overall density for a piece of gas that stays in the same relative position in the cloud will then follow
\eq{toy:nadiab}{
n_c^{\rm adiabatic}(t) = n(0)e^{-3t/\tau_{\rm release}}.
}
To incorporate this effect into the model it is combined (somewhat ad hoc) with the expression $n_c(r_{c0}(t),t)$ for instantaneous release to give the final estimate of local density for the $k$ momentum escaping atoms used in \eqn{eq:toy} as
\eq{toy:nt}{
n(t) = n(0)e^{-3t/\tau_{\rm release}} + (1-e^{-3t/\tau_{\rm release}})\ n_c( r_{c0}(t),t).
}
with the help of \eqn{toy:decay} and \eqn{toy:rc0}.

\clearpage
\renewcommand{\thepage}{\arabic{page}}
\setcounter{page}{19}

\end{document}